 \documentclass[10pt,twocolumn,preprintnumbers,amsmath,amssymb]{revtex4}

 \usepackage{color}
 \usepackage{docs}
 \usepackage{bm}
 \usepackage{graphicx}
 \usepackage{latexsym}
 \usepackage{pifont}

 \newcommand{\Fig}[1]{Fig.~\ref{fig:#1}}
 \newcommand{\Tab}[1]{Table.~(\ref{tab:#1})}
 \newcommand{\Eqn}[1]{Eq.~(\ref{eq:#1})}
 \newcommand{\Sec}[1]{Sec.~\ref{sec:#1}}

 \newcommand{\iotabar}{\mbox{$\iota\!\!$-}}

 \newcommand{\boldxi}{\mbox{\boldmath$\xi$\unboldmath}}
 \newcommand{\boldkappa}{\mbox{\boldmath$\kappa$\unboldmath}}
 \newcommand{\boldlambda}{\mbox{\boldmath$\lambda$\unboldmath}}

 \newcommand{\s}{s}
 \renewcommand{\t}{\theta}
 \newcommand{\z}{\zeta}

 \newcommand{\intds}[2]{\int_{s_{#1}}^{s_{#2}} \!\!\!\!\!\!\!\! ds \,}
 \newcommand{\intdt}{\int_{0}^{2\pi} \!\!\!\!\!\!\! d\theta}
 \newcommand{\intdz}{\int_{0}^{2\pi} \!\!\!\!\!\!\! d\zeta}

 \newcommand{\be}{\begin{eqnarray}}
 \newcommand{\ee}{\end{eqnarray}}

 \newcommand{\Poincare}{Poincar$\acute{\rm e}$ }
 \newcommand{\PoincareBirkhoff}{Poincar$\acute{\rm e}$-Birkhoff }
 \newcommand{\Schluter}{Schl$\ddot{\rm u}$ter }

 \newcommand{\insertsglfigure}[3]{
  \begin{figure}[t]
  \begin{center}
  \includegraphics{#1}
  \vspace{0.0cm}
  \caption{#2 \label{fig:#3}}
  \end{center}
  \end{figure}
 }

 \newcommand{\insertdblfigure}[3]{
  \begin{figure*}[th]
  \begin{center}
  \includegraphics{#1}
  \vspace{0.0cm}
  \caption{#2\label{fig:#3}}
  \end{center}
  \end{figure*}
 }

 \renewcommand{\vec}[1]{\mbox{$\bf #1$}}

 \newcommand{\Id}{\textsf{\textbf{I}}} 

 \newcommand{\const}{\mbox{const}}

 \newcommand{\dotv}{ \cdot }

 \newcommand{\ds}{\displaystyle}

 \newcommand{\grad}{\nabla}
 
 \newcommand{\divv}{ \grad\dotv }
 \newcommand{\curl}{\nabla\times}

 \renewcommand{\d}{ { d} } 

\begin{document}

 \title{Computation of multi-region relaxed magnetohydrodynamic equilibria}

 \author{S.R. Hudson}
 \email{shudson@pppl.gov}

 
 \author{R.L. Dewar}
 \email{robert.dewar@anu.edu.au}

 \author{G. Dennis}
 \email{graham.dennis@anu.edu.au}

 \author{M.J. Hole}
 \email{matthew.hole@anu.edu.au}

 \author{M. McGann}
 \email{mathew.mcgann@anu.edu.au}

 \author{G. von Nessi}
 \email{greg.vonnessi@anu.edu.au}


 \author{S. Lazerson}
 \email{slazerso@pppl.gov}

 \affiliation{Princeton Plasma Physics Laboratory, PO Box 451, Princeton NJ 08543, USA}
 \affiliation{Plasma Research Laboratory, Research School of Physics \& Engineering, The Australian National University, Canberra, ACT 0200, Australia}

 \date{\today}

 \begin{abstract}
 
 We describe the construction of stepped-pressure equilibria as extrema of a multi-region, relaxed magnetohydrodynamic (MHD) energy functional that combines elements of ideal MHD and Taylor relaxation, and which we call MRXMHD.
 The model is compatible with Hamiltonian chaos theory and allows the three-dimensional MHD equilibrium problem to be formulated in a well-posed manner suitable for computation.
 The energy-functional is discretized using a mixed finite-element, Fourier representation for the magnetic vector potential and the equilibrium geometry; and numerical solutions are constructed using the stepped-pressure equilibrium code, SPEC.
 Convergence studies with respect to radial and Fourier resolution are presented.

 \end{abstract}

 \maketitle

 \section{Introduction} \label{sec:introduction}
 
  Zero-Larmor-radius, single-fluid magnetohydrodynamics (MHD) is commonly used for modeling the global, long-time-scale state of plasmas in the magnetic confinement devices used for fusion power research.
  It is often reasonable to approximate the plasma pressure tensor as isotropic and to ignore inertial effects due to small mass flows.
  There is no minimum length scale in this model, so spatial discontinuities are allowed \cite{MHDvN_10}.
  To allow a weak formulation, we write the equilibrium condition in conservation form,
 \be \label{eq:stressbalance}
 \divv \left( p \, \Id + \frac{B^2}{2\mu_0} \Id - \frac{\vec{B}\vec{B}} {\mu_0} \right) = 0,
 \ee
 where, using SI units, $\mu_0$ is the permeability of free space, $p(\vec{r}) \geq 0$ is the pressure as a function of position \mbox{$\vec{r} = x\vec{i} + y\vec{j} + z\vec{k}$}, and $\vec{B}(\vec{r})$ is the magnetic field, which must obey $\divv\vec{B} = 0$.
  While MHD is a rather crude model for the physics of a plasma, the Maxwell equations for the magnetic field and the `dynamics' of field lines are exact.

  The problem addressed in this paper is, treating both $p$ and $\vec{B}$ as unknown fields within suitable function spaces, find general, weak solutions of \Eqn{stressbalance} in an arbitrary, three-dimensional (3D) toroidal domain, ${\cal V}$, under the homogeneous boundary conditions,
 \be \label{eq:pBbcs} 
  p = 0, \; \vec{n}\dotv\vec{B} = 0, \; \forall \: \vec{r} \in {\partial \cal V},
 \ee
 where $\vec{n}$ is the unit normal at the boundary, $\partial {\cal V}$.
  We take the boundary to be fixed, being either the edge of a plasma confined by a notional, tight-fitting shell, or the boundary of a surrounding vacuum region.

  Our goal is to formulate the 3D equilibrium problem in a way that is as well-posed mathematically as the two-dimensional (2D) problem, by which we mean that a well-defined, unique solution exists.
  And, to develop an accurate, robust and efficient numerical solution method, where the error between the approximate numerical solution, e.g. $f_h$, and the exact solution, $f$, is bounded and goes to zero as $f_h = f + {\cal O}(h^n)$, where $h$ characterizes the numerical resolution and $n$ depends on the numerical discretization.
 
  If $p$ and ${\bf B}$ are assumed to be differentiable within a subregion of $\cal V$, then \Eqn{stressbalance} is locally equivalent to the force-balance condition
 \be \label{eq:forcebalance}
  \grad p = \vec{j}\times\vec{B},
 \ee
 where $\vec{j} = \curl\vec{B}$ is the current density (here, and hereafter, $\mu_0$ is ignored).
  We will {\em not} restrict attention to differentiable solutions in the following, but we will work within the approximation that ${\bf B}\cdot \nabla p = 0$, which follows directly from \Eqn{forcebalance}.
  Physically, this approximates the transport of heat and mass along the magnetic field as infinite compared to that across the field.
  This immediately implies that $p$ is invariant along the magnetic field: the spatial dependence of the pressure and the phase space structure of the magnetic field are intimately connected.

  A specific equilibrium state is characterized by the pressure, i.e. $p$ is considered to be a supplied, input function.
  The computational challenge is to then determine the magnetic field that is consistent with the given pressure and boundary.
  Generically, in 3D, there exist regions within ${\cal V}$ where the magnetic field lines are chaotic.
  To admit numerically tractable solutions for ${\bf B}$, it is necessary to restrict the class of admissible functions for $p$; and to guarantee that ${\bf B}$ is consistent with a given $p$, topological constraints on ${\bf B}$ must be enforced.

  In \Sec{hamiltonianchaos}, we review the salient properties of 3D magnetic fields, which generally have a fractal phase space, and we sketch the nature of continuous solutions for $p$ and $\vec{B}$.
  This is based on the construction of an ergodic partition; which, being fractal, is impractical from a standpoint of numerical implementation.
  So, we describe a discrete partition which greatly simplifies the equilibrium problem and leads naturally to stepped-pressure equilibria, where the plasma is modeled as a set of nested volumes in each of which the field satisfies the Beltrami equation, $\nabla \times {\bf B} = \mu {\bf B}$, and across the interfaces that separate these volumes the total pressure is continuous, $[[p+B^2/2]]=0$.

  `Sharp-boundary' \cite{BFLMT_86} states and multi-volume \cite{DHMMH_08} sharp-boundary states have been considered previously, and Bruno \& Laurence \cite{Bruno_Laurence_96} have presented theorems that insure the existence of sharp boundary solutions, with an arbitrary number of pressure jumps, for tori whose departure from axisymmetry is sufficiently small.
  In \Sec{mrxmhd} we introduce a variational approach to solving \Eqn{stressbalance} based on the notion of multi-region, relaxed MHD (MRXMHD), which is a generalization of Taylor's \cite{Taylor_86} relaxed-MHD formulation: that a sufficiently turbulent/chaotic, weakly non-ideal plasma will evolve so as to minimize the energy subject to the constraint of conserved magnetic helicity, and in doing so will break most of the constraints \cite{Yoshida_Dewar_12} of ideal MHD, thus allowing magnetic reconnection.
  In MRXMHD, a plasma with a non-trivial pressure profile is constructed as a nested collection of relaxed states, between which the ideal-MHD constraints apply.
  By deriving the Euler-Lagrange equations, we see that the MRXMHD energy functional has stepped-pressure equilibria as extremizing solutions.
  A close examination of the force-balance condition, $[[p+B^2/2]]=0$, reveals that the rotational transform of the interfaces must be strongly irrational.
 
  In \Sec{discretizedenergyfunctional}, the MRXMHD energy functional is discretized using a mixed Fourier, finite-element representation for the vector potential and geometry.
  Setting to zero the derivatives of the energy functional with respect to the vector-potential in each volume gives a linear system for the magnetic field, $\nabla\times{\bf B}=\mu {\bf B}$, where $\mu$ is a Lagrange multiplier (sometimes called the Beltrami parameter).
  This can be adjusted in order to preserve the helicity integral, or both $\mu$ and the enclosed poloidal flux can be adjusted to satisfy the interface rotational transform constraints.

  Assuming the Beltrami fields in each volume have been computed for a arbitrary interface geometry, the problem of constructing an equilibrium solution is standard: changes in the interface geometry are allowed to either minimize the energy functional using conjugate gradient methods, or to find a zero of the multi-dimensional gradient $\equiv$ force-balance vector using a Newton method.
  To fully constrain the Fourier representation of the interface geometry, we employ spectral-condensation \cite{Hirshman_Meier_85,Hirshman_Breslau_98} methods to obtain a preferred poloidal angle coordinate.
  Illustration of equilibrium states and convergence studies are then presented.

  At appropriate points in the discourse, we contrast our approach to constructing equilibrium solutions with others in the literature.

 \section{Hamiltonian chaos, partitioned} \label{sec:hamiltonianchaos}

  Magnetic-field-line flow is a Hamiltonian system \cite{Boozer_RMP}.
  The well-developed theory of Hamiltonian dynamical systems (see, for example, the texts by Wiggins \cite{Wiggins_90} and Lichtenberg \& Lieberman \cite{Lichtenberg_Lieberman_92}, and the review by Meiss \cite{Meiss_92}) provides a strong foundation on which to build.
  We shall sometimes use general dynamical-systems language rather than the more specialized plasma terminology; for instance, using `orbit' and `magnetic field line' interchangeably.
  To facilitate the following discussion, we use cylindrical coordinates, $(R,\phi,Z)$, which are orthogonal and right handed, so that $x=R\cos(\phi)$, $y=R\sin(\phi)$, and $z=Z$, and $(x,y,z)$ are Cartesian.

  Devices of the tokamak and reversed-field-pinch (RFP) \cite{Boozer_RMP} classes use a large number of identical toroidal field coils arranged with a discrete rotational symmetry about the $z$-axis.
  In the axisymmetric special case, it is reasonable to seek solutions that are invariant under rotation.

  Axisymmetric magnetic fields are representable as 1-degree-of-freedom (1-dof) autonomous Hamiltonian systems \cite{Boozer_RMP}, with $\phi$, periodic, playing the role of time.
  Such systems are integrable in the dynamical systems sense, and action-angle coordinates may be constructed.
  The field lines lie on nested invariant tori, $\psi = \const$, that foliate the extended phase space, $(\psi,\t,\phi)$, where $\psi$ is a toroidal flux function and $\t$ is a poloidal angle that increases linearly against $\phi$.
  In the terminology of magnetic confinement, the invariant tori are called magnetic flux surfaces, and action-angle coordinates are called straight-field-line coordinates.
  In the following, when we refer to an integrable system, we will assume that the integrable system has shear.

  The invariant tori $\equiv$ flux surfaces are characterized by their rotation number $\equiv$ rotational transform, which is commonly denoted in magnetic confinement plasma physics by $\iotabar$.
  (Historically \cite{Kruskal_Kulsrud_58}, the term rotational transform refers to $\iota$, the average poloidal angle increase in each iteration of the return map, but in modern usage \cite{Boozer_RMP} it is used for $\iotabar \equiv \iota/2\pi$, and often the `bar' is omitted.)
  If $\iotabar$ is a rational number, $\iotabar=n/m$ where $n$ and $m$ are integers, then the corresponding surface is foliated by periodic orbits $\equiv$ closed field lines, which close on themselves after $m$ toroidal transits, having undergone $n$ poloidal transits.
  If $\iotabar$ is an irrational number, then the flux surface is covered ergodically by a single {\em quasi-periodic} orbit, which never closes on itself (but comes arbitrarily close), and each irrational surface is the closure of an irrational field line.

  In the axisymmetric case, the equilibrium problem can be reduced to the task of solving a 2D partial differential equation, the Grad-Shafranov equation \cite{Boozer_RMP,NCGL_08}, which is well-posed (except for bifurcations \cite{Solano_04}).
  The equilibria are characterized by two free profile functions, e.g. the pressure, $p(\psi)$, and the rotational transform, $\iotabar(\psi)$.
  Because space is foliated by flux surfaces, equilibria with continuous, smooth profiles are admissible; in fact, the only magnetic fields consistent with ${\bf B}\cdot\nabla p = 0$ and globally smooth profiles are integrable magnetic fields.

  Axisymmetry is necessarily always broken to some extent by the modular nature of the conductors and machine imperfection, or by intentionally applied perturbation fields \cite{Evans_Moyer_Thomas_etal_04}, or by equilibrium bifurcations \cite{CGPSV_10}.
  The {\em stellarator} family \cite{Spitzer_58,Boozer_RMP} of confinement devices is intentionally nonaxisymmetric.
  This allows greater freedom in the design of experiments and can provide enhanced plasma stability.
  (The nonaxisymmetry of stellarators, however, generally leads to degraded particle confinement; this can be ameliorated, somewhat, by the use of `quasi-symmetric' configurations \cite{Nuehrenberg_Zille_88}.)

  The 3D magnetic field-line flow is still analogous to a Hamiltonian dynamical system, but because there is no longer a symmetry coordinate, the 3D field-line Hamiltonian is not autonomous.
  Such systems, still {\em periodic} in $\phi$, are called $1\frac{1}{2}$-dof systems and are generically non-integrable, meaning that the extended phase space is almost {\em never} foliated by invariant tori.

  The periodic orbits are fragile.
  Resonant magnetic fields associated with geometric deformation destroy almost all of the periodic orbits; magnetic islands form, and regions of chaotic magnetic field lines emerge.
  The destruction of rational surfaces is related to the classical problem of small denominators in the transformation to action-angle coordinates for the perturbed system.
  The \PoincareBirkhoff theorem \cite{Meiss_92} shows that, for every rational invariant torus present in the integrable case, at least two of the periodic orbits survive.
  One orbit is hyperbolically unstable, while the other is elliptically stable or has become hyperbolic through a period-doubling bifurcation.

  These orbits, known as \PoincareBirkhoff orbits, form a robust `skeleton' of invariant sets and provide crucial information about the structure of phase space. 
  The existence of a given KAM surface (described below) can be inferred from the stability of nearby periodic orbits using Greene's residue criterion \cite{Greene_79}.
  Associated with each unstable periodic orbit is an unstable manifold and a {\em chaotic sea} \cite{Wiggins_90}, comprised of {\em irregular} trajectories without a well defined rotational transform, i.e. the ratio $\Delta \t / \Delta \phi$ does not converge as $\Delta \phi \rightarrow \infty$, where $\Delta \t$ and $\Delta \phi$ are the increase in $\t$ and $\phi$ along a field line.
  Although there is no formal proof \cite{Meiss_92}, it is standard to assume, based on computational evidence, that the closure of each chaotic sea is a three-dimensional subset of $\mathbb{R}^3$, as each irregular trajectory seems to fill a volume.
  Associated with the elliptic periodic orbits are local regions of regular trajectories, the so-called magnetic islands.

  The irrational field lines are quite robust to perturbation.
  Indeed, they are guaranteed to survive by the Aubry-Mather theorem; however, a given irrational field line may or may not come arbitrarily close to every point on a smooth surface.
  The KAM theorem, named in honor of Kolmogorov, Arnold and Moser \cite{Kolmogorov_54,Moser_62,Arnold_63,Moser_73,Arnold_78}, shows that a finite measure of invariant tori {\em do} exist for sufficiently small, smooth perturbations to an integrable system, provided that the rotational transform, $\iotabar$, is sufficiently irrational, i.e. $\iotabar$ must satisfy a {\em Diophantine} condition: there exists an \mbox{$r>0$} and \mbox{$k\ge 2$} such that, for all integers \mbox{$n$} and \mbox{$m$}, \mbox{$| \iotabar - n/m | > r/m^k$}.
  About each rational, $n/m$, there is an excluded region of width $r/m^k$, which is consistent with the emergence of a chaotic sea about every unstable periodic orbit.
  KAM tori are two-dimensional subsets of $\mathbb{R}^3$ whose union is of finite measure and forms a partition of phase space.

  Typically, as the magnitude of the geometric deformation increases, the size of the magnetic islands increases, the volume of the chaotic seas increases, and each given KAM surface will become more geometrically deformed until a critical point is reached, at which point the surface is continuous but no longer smooth.
  These critical tori form fractal boundaries between the chaotic seas associated with different island chains.
  By `fractal' we simply mean having a hierarchy of qualitatively self-similar structure on all scales, with no minimum length scale, and is non-differentiable.

  Some KAM tori are more robust than others.
  The most robust invariant tori are those that have the `most irrational' rotational transforms, where `most irrational' means most difficult to approximate with rationals.
  Such irrationals are called noble, and their definition is made precise using the continued fraction representation \cite{Niven_56}.
  The noble KAM tori are also the smoothest, in that fewer Fourier harmonics are required for an accurate description of their geometry.

  After the destruction of a KAM surface, the closure of an irrational field line has the structure of a Cantor set \cite{Aubry_83,Mather_82} and is called a {\em cantorus} \cite{Percival_79a} (hint: Cantor + torus = cantorus).
  Cantori are one-dimensional subsets of $\mathbb{R}^3$ \cite{Li_Bak_86}, and constitute a set of zero measure that does not serve to partition phase space.
  The cantori can, however, form effective {\em partial} barriers to field-line transport \cite{MacKay_Meiss_Percival_84_PRL} and thus also to anisotropic diffusion \cite{Hudson_Breslau_08}.
 
  Ergodic invariant sets form a fractal hierarchy.
  The `primary' chaotic seas and KAM tori are infinitely intertwined, and each chaotic sea contains `secondary' island chains, KAM tori, cantori, and chaotic seas in an `islands around islands' pattern repeated {\em ad infinitum} \cite{Mackay_Meiss_Percival_84,Meiss_86,Meiss_92,Mezic_Wiggins_99,Levnajic_Mezic_10}.
  The chaotic seas are infinitely multiply-connected, bounded externally by critical, primary invariant tori, and internally by the infinite hierarchy of islands.

 \subsection{continuous solution on ergodic partition} \label{sec:generalergodicsolution}

  To understand the general class of functions for $p$ and ${\bf B}$ that admit solutions to the equilibrium problem we first consider the implications that the non-integrability of the magnetic field has on the structure of the pressure; and then, given a pressure that is consistent with a non-integrable field, consider the implications this has on the field itself.

  To understand the structure of the pressure function that satisfies ${\bf B} \cdot \grad p = 0$, given a generic magnetic field, it is convenient to represent phase space as a collection of pair-wise disjoint sets that are invariant under the field-line flow map, $\bm{\varphi}_{\phi}:\vec{r}_0 \mapsto \vec{r}$.
  This map is constructed simply by following a field line a distance $\phi$ in toroidal angle from a point, $\vec{r}_0$, on a surface of section (for example the $\phi = 0$ plane) to arrive at point $\vec{r}$.
  The {\em return map} is generated by following field lines once around the machine back to the initial surface of section.
  (In the case of the RFP device, this discussion applies in a subdomain not containing points where $B^{\phi}$ reverses sign -- to treat the toroidal field-reversal region a poloidal surface of section should be used instead.)
  An invariant set \mbox{${\cal A} \subset \mathbb{R}^2$} within a surface of section is a set invariant under the return map, \mbox{$\bm{\varphi}_{2\pi}({\cal A}) = {\cal A}$}.
  An invariant set \mbox{${\cal V} \subset \mathbb{R}^3$} within phase space may be constructed as the continuous union of such sets, i.e. \mbox{${\cal V} = \bigcup_{\phi \in [0,2\pi)}\bm{\varphi}_{\phi}({\cal A})$}.

  An invariant partition is a union of invariant {\em tori}, which are two-dimensional magnetic surfaces, and three-dimensional invariant toroidal volumes or {\em toroids} bounded by invariant tori.
  If an invariant volume contains an invariant surface, e.g. a KAM surface, then the volume may be subdivided into two distinct subvolumes, each of which is invariant under the return map.
  An {\em ergodic} invariant set is a set with finite measure that allows no further subdivision, 
 and the {\em ergodic partition} of phase space is its decomposition into ergodic invariant sets and a nonergodic (periodic) set of zero measure -- see definition 2.1 of Ref.\cite{Mezic_Wiggins_99}.

  For the purpose of constructing weak solutions to the equilibrium problem, we are primarily interested in the sets of finite measure.
  We ignore the cantori and periodic orbits and take our partition as having every chaotic sea, ${\cal C}_{\alpha}$, each of which has finite volume, and the invariant surfaces, ${\cal C}_{\beta}$, the union of which has finite measure, as the only elements with non-trivial measure, where $\alpha$ and $\beta$ are elements of appropriate indexing sets (e.g. $\alpha$ is a rational and $\beta$ is irrational).
  As any field line approaches every point arbitrarily closely in a given ergodic set, the only solution for $p$ consistent with ${\bf B}\cdot \nabla p = 0$ is 
 $p = p_{\alpha} = \const$, $ \forall \, \vec{r} \in {\cal C}_{\alpha}$, and similarly for ${\cal C}_\beta$.
  The most general solution for the pressure is
 \be \label{eq:pgensol}
  p(\vec{r}) = p_\alpha I_\alpha(\vec{r}) + p_\beta I_\beta(\vec{r}),
 \ee
 where $I_\alpha$ is an indicator function on each ergodic component, i.e. $I_\alpha(\vec{r}) = 1$ if $ \vec{r} \in {\cal C}_{\alpha}$ and $I_\alpha(\vec{r}) = 0$ otherwise, and similarly for $I_\beta$.

  We now recognize that $\vec{B}(\vec{r})$ is {\em not} arbitrary and seek a similarly general characterization of the constraints that \Eqn{forcebalance} places on this function.
  Each ${\cal C}_{\alpha}$ has finite volume, and we assume that ${\bf B}$ is differentiable within ${\cal C}_\alpha$.
  That the pressure is constant in ${\cal C}_{\alpha}$ implies that $\nabla p=0$.
  Then, force balance, $\nabla p = {\bf j}\times{\bf B}$, implies that $\curl\vec{B} = \mu(\vec{r})\vec{B}$, for some scalar function $\mu({\bf r})$.
  Taking the divergence of this equation, we find $\vec{B} \dotv\grad \mu = 0$.
  Thus, like $p$, $\mu$ must be constant within each ergodic region \cite{Rusbridge_77}, $\mu=\mu_\alpha$ in ${\cal C}_{\alpha}$, and $\vec{B}$ must be a linear force-free field, i.e. it satisfies the Beltrami equation,
 \be \label{eq:Beltrami}
  \curl\vec{B} = \mu_{\alpha} \vec{B}.
 \ee
 
  This is a well-studied linear elliptic partial differential equation, about which much is known \cite{Kress_81,Kress_86,Yoshida_Giga_90,Yoshida_92,Marsh_96,Hudson_Hole_Dewar_07}.
  To construct a solution in a given domain it is required to specify (i) the boundary of the domain; (ii) appropriate boundary conditions, e.g. $\vec{n}\dotv\vec{B} = 0$, where $\vec{n}$ is the unit normal; and (iii) homological conditions, i.e. line integrals (fluxes) around topologically inequivalent loops.
  To specify a solution in a simple torus it is sufficient to specify $\mu$ and the toroidal flux, while in a doubly-connected annulus the poloidal flux must also be specified \cite{Hudson_Hole_Dewar_07}.

  Solving the Beltrami equation in general ${\cal C}_{\alpha}$ is, however, an intractable numerical problem.
  Because of the topological complexity resulting from the infinity of islands embedded in the chaotic sea, there is an infinity of inequivalent closed loops.
  The outer boundary of each chaotic sea is presumably a critical KAM torus, which is {\em not} smooth, and the normal to the fractal boundary is not defined.

  Furthermore, a continuous, non-trivial pressure, that is consistent with a generic non-integrable field, must be fractal.
  To see this, we may assume that a finite pressure gradient is supported by the KAM tori.
  The Diophantine condition serves as a simple, proxy indicator function describing the existence of KAM tori in the fractal phase space of a generic non-integrable field (though the more complicated Bruno function \cite{Locatelli_etal_00} is probably a better approximation).
  Let us consider a {\em Diophantine} pressure profile, $p(\iotabar)$, defined $p'(\iotabar) = 1$ if \mbox{$| \iotabar - n/m | > r/m^k$} for all integers \mbox{$n$} and \mbox{$m$}, and $p'(\iotabar)=0$ otherwise, supplemented with the condition $p(0)=0$.
  The function $p(\iotabar)$ is continuous by construction (i.e. the derivative is nowhere infinite) and we assume that $r$ and $k$ have been chosen so that $p'(\iotabar)$ is non-zero on a set of finite measure (so that not all the excluded regions overlap) so that $p(\iotabar)$ is non-trivial.
  Even for this `toy' model, numerically approximating the function $p(\iotabar)$ given $p'(\iotabar)$ is rather complicated.
 
  An approximation to $p(\iotabar)$ may (in the case of continuous $p'$) be constructed using a tagged partition, i.e. $p(\iotabar) \approx \sum_{i}p'(x_i)(\iotabar_{i}-\iotabar_{i-1})$, where $x_i \in [\iotabar_{i-1},\iotabar_{i}]$ and $0=\iotabar_0 < \iotabar_1 < \dots < \iotabar_N = \iotabar$.
  However, because $p'(\iotabar)$ is nowhere continuous except where $p'(\iotabar)=0$, the result depends on the choice of $x_i$ even when $|\iotabar_{i}-\iotabar_{i-1}| \rightarrow 0, \forall i$.
  That is, the Riemann integral of $p'(\iotabar)$ does not exist.
  The error between this approximation and the exact solution is not bounded.

  More sophisticated numerical discretizations could be derived; for example, by choosing the $x_i$ in the tagged partition to coincide with the locally most irrational, i.e. by constructing an `irrational' tagged partition \cite{Hudson_04}.
  However, a precise treatment would involve numerically approximating the Lebesgue integral and complicated measure theory.
  In 1967, Grad \cite{Grad_67} made a similar comment, describing the pressure as ``pathological''.

  Furthermore, the fractal structure of non-integrable fields in toroidal confinement devices will be far more complicated than that described by the Diophantine condition, and may not (will not!) be known apriori.
  There are numerical diagnostics for determining the structure of phase space, such as Greene's residue criterion, but these diagnostics come at considerable computational cost.

  Considering
  (i) that a nonlinear equilibrium calculation will inevitably require an iterative approach, in which the fractal phase space structure of the field may need to be re-evaluated at each iteration, and
  (ii) that the critical KAM tori are fragile, and an infinitesimal change in ${\bf B}$ can cause an abrupt, finite change in the volume of any given chaotic sea, and 
  (iii) that the fractal structure of phase space will need to be resolved sufficiently accurately in order to guarantee that an appropriately defined error is below some bound, that can be made arbitrarily small as the numerical resolution is increased;
  we may expect that this computational cost would be excessive.

  For our purpose of constructing a robust and efficient numerical solution of well-defined, 3D MHD equilibria with non-integrable fields, with a bounded error that can be made arbitrarily small, it is far better to work with smooth functions, and to employ an algorithm that does not depend on resolving the infinitely complicated structure of phase space.
  So, we extend our class of functions for $p$ and ${\bf B}$ beyond {\em globally} continuous functions, as non-trivial, continuous functions that satisfy force balance are necessarily {\em fractal}, and consider instead functions that are continuous, and smooth, {\em almost} everywhere; that is, we consider functions that are smooth except for a finite set of discontinuities, which can be easily managed numerically.

 \subsection{weak solution on discrete partition} \label{sec:simplifiedpartition}

  Continuous pressure profiles are not the most general solutions of \Eqn{stressbalance}.
  Discontinuous pressure profiles may seem unphysical, but they are a valid solution class within the zero-Larmor-radius MHD model we have adopted.
  If continuous, globally smooth solutions are required, then additional `non-ideal' physics should be included \cite{PMSM_86}.
  For example, including a small, but finite, diffusion of the pressure perpendicular to the magnetic field will provide solutions with a globally smooth pressure; and including a small resistivity will prevent the formation of singular currents.
  Appropriate source terms are required to balance dissipative effects.

  This is the approach adopted by various codes \cite{SPECYL,PIXIE3D,M3DC1,NIMROD} that can approximate an MHD equilibrium as a resistive steady state, but which are best described as initial-value, time-evolution codes and cannot, strictly, compute an equilibrium that satisfies ${\bf B}\cdot \nabla p=0$, with the pressure given.
  The algorithms these codes employ become increasingly ill-conditioned as the non-ideal terms approach zero \cite{dCNC_11}.

  We now describe a restriction of the solution class that greatly simplifies the equilibrium problem.
  A {\em discrete} invariant partition of phase space is constructed.
  The disjoint, invariant sets that are surrounded by a given primary chaotic sea, e.g. the hierarchy of island chains, are absorbed into the chaotic sea itself, which then becomes either a simply or doubly connected region; and the outer boundary of these regions is extended past the adjacent, critical boundary surface to a smooth, noble surface, which also serves as the boundary for the next `extended' chaotic sea.
  That is, we choose a set of smooth, noble KAM tori, ${\cal I}_l$, where $l = 1,2,\ldots N_V$, that partitions phase space into $N_V$ invariant toroidal or annular subvolumes, ${\cal V}_l$.
  Each ${\cal V}_l$ is an invariant set under the field line map, but not necessarily an ergodic invariant set because the field may not be totally chaotic.

  In each region, ${\cal V}_{l}$, we equate all the $\mu_{\alpha'}$ to a single constant $\mu_l$, and all the $p_{\alpha'}$ and $p_{\beta'}$ to a single constant $p_l$, where $\alpha'$ and $\beta'$ label all the chaotic seas and invariant tori within ${\cal V}_{l}$.
  Each ${\cal V}_{l}$ is simply or doubly connected with a smooth boundary and it is a simple computational task to solve $\nabla \times {\bf B}_l = \mu_l {\bf B}_l$ in each ${\cal V}_l$.
  We will enforce the constraint that ${\bf n}\cdot{\bf B}=0$ on the ${\cal I}_l$, but otherwise the topology of the field in each ${\cal V}_l$ is unconstrained.

  For the pressure, rather than restricting attention to a {\em globally} continuous pressure with finite pressure-{\em gradient} on the uncountably infinite ${\cal C}_{\beta}$, we instead consider a {\em piecewise} continuous pressure with finite pressure-{\em jumps} on the finite set ${\cal I}_{l}$.
  Intuitively, we imagine that all of the pressure in the {\em continuous-but-fractal} model that is supported by the ${\cal C}_{\beta}$ in the vicinity of a selected noble KAM torus is placed on the noble torus itself: all of the pressure is placed on a finite selection of the most irrational surfaces.
  The vanishing of the divergence of the stress tensor, \Eqn{stressbalance}, in a neighborhood of a surface of discontinuity gives a condition \cite{MHDvN_10} that must be satisfied at the interfaces, namely that the total pressure must be continuous across the ${\cal I}_l$, i.e. $[[p+B^2/2]]=0$.

  We have described the interfaces where there exists a discontinuity in the pressure and the tangential field as KAM surfaces, but this is rather loose terminology.
  Such interfaces are perhaps `double-sided' KAM surfaces, being covered by an field line $\equiv$ integral curve with irrational frequency, of the field, ${\bf B}_-$, immediately inside the torus, while also being covered by an integral curve with the same irrational frequency of the perhaps different field, ${\bf B}_+$, immediately outside the torus.
  In the next section, where we describe the MRXMHD energy functional, we shall refer to  the ${\cal I}_l$ as {\em ideal interfaces} and the ${\cal V}_l$ as {\em relaxed volumes}, and describe why the ${\cal I}_l$ are required to have irrational rotational transform.

  In the above discussion, we have argued that {\em stepped-pressure equilibria} arise naturally when one seeks a numerically tractable discretization of the equilibrium problem that is consistent with the zero-Larmor-radius model of MHD; satisfies ${\bf B}\cdot \nabla p = 0$; and is consistent with what is known about the fractal phase space structure of non-integrable fields.
  The equilibria could also be described as multi-volume sharp-boundary states.
  A  major motivation for pursuing this model is that Bruno \& Laurence \cite{Bruno_Laurence_96} have proven that such stepped-pressure equilibria exist (provided the departure from axisymmetry is sufficiently small).
  The number of volumes, $N_V$, and interfaces may be made arbitrarily large.
  We can, depending on the numerical resources available, consider a sequence of invariant partitions with increasing $N_V$, so the discontinuities in the pressure are made arbitrarily small, in order to study the nature of a continuous-but-fractal equilibrium via a sequence of well-defined, stepped-pressure equilibria.
  Stepped-pressure profiles are sufficiently general to represent observed profiles to within experimental error.

  In order to explore the properties of these equilibria in arbitrary geometry, i.e. to go beyond what may be proved analytically, this model has been implemented numerically in the stepped-pressure equilibrium code, SPEC, as will be described below in \Sec{discretizedenergyfunctional}.
  We now show that there is an multi-region, relaxed MHD energy functional, that we call MRXMHD, that has stepped-pressure equilibria as extremizing solutions.

 \section{energy functional method} \label{sec:mrxmhd}

  The classic MHD energy functional \cite{Kruskal_Kulsrud_58} is given by the integral
 \be \label{eq:classicenergyfunctional}
  W \equiv \int_{{\cal V}} \left( \frac{p}{\gamma-1} + \frac{B^2}{2} \right) dv,
 \ee
 where ${\cal V}$ is the plasma volume bounded by a toroidal surface, $\partial{\cal V}$.
  Ideal equilibria are obtained when the plasma is in a minimum energy state: more precisely, when the energy functional is extremized allowing for a restricted class of variations, namely ideal variations.
  The equation of state, \mbox{$d_t(p/\rho^\gamma)=0$}, where \mbox{$d_t \equiv \partial_t + {\bf v} \cdot \nabla$} and ${\bf v}$ is the `velocity' of an assumed plasma displacement, \mbox{${\bf v}=\partial_t \boldxi$}, may be combined with mass conservation, \mbox{$\partial_t \rho + \nabla \cdot ( \rho \, {\bf v} ) = 0$}, to obtain an equation that constrains the variation in the pressure, \mbox{$\delta p = (\gamma-1) \boldxi \cdot \nabla p - \gamma \nabla \cdot ( p \, \boldxi )$}.
  Faraday's law, \mbox{$\partial_t {\bf B} = \nabla \times {\bf E}$}, may be combined with the ideal Ohm's law, \mbox{${\bf E}+{\bf v}\times{\bf B}=0$}, where \mbox{${\bf E}$} is the electric field, to obtain an equation that constrains the variation in the magnetic field, $\delta {\bf B} = \nabla \times ( \boldxi \times {\bf B} )$.
  Note that this last constraint does not allow the topology of the field to change.

  The first variation in the energy due to an ideal displacement, $\boldxi$, that is assumed to vanish on the boundary, is given by
 \be \label{eq:classicenergyfunctionalidealvariation}
  \delta W \equiv \int_{{\cal V}} ( \nabla p - {\bf j} \times {\bf B}) \cdot \boldxi \, dv.
 \ee
  Extremizing solutions satisfy the ideal force-balance condition, $\nabla p = {\bf j}\times{\bf B}$.
  In order to uniquely define an equilibrium, in addition to the shape of the plasma boundary, it is required to specify the pressure, and either the rotational-transform or the parallel current density \cite{Bauer_Betancourt_Garabedian_84,Boozer_RMP}.
  (Note that the constraints of ideal MHD places a constraint on the differential toroidal and poloidal fluxes and the rotational-transform, namely $d\psi_p / d\psi_t = \iotabar$.)

  The VMEC \cite{Hirshman_Whitson_83} and BETAS/NSTAB codes \cite{Betancourt_88,Taylor_94} are based on this approach.
  These codes assume that the magnetic field is integrable and allow for smooth pressure and rotational-transform profiles.

  In general 3D geometry, there is a singularity in the resonant harmonic of the parallel current in equilibria with nested flux surfaces \cite{BHHNS_95}.
  Writing the current as \mbox{${\bf j} = \sigma {\bf B} + {\bf j}_\perp$}, the quasineutrality condition, \mbox{$\nabla \cdot {\bf j}=0$}, requires that the parallel current must satisfy the magnetic differential equation, \mbox{${\bf B} \cdot \nabla \sigma = -\nabla \cdot {\bf j}_\perp$}, where we may consider the perpendicular current to be driven by the pressure gradient, \mbox{${\bf j}_\perp = {\bf B} \times \nabla p / B^2$}.
  Magnetic differential equations are densely singular \cite{Newcomb_59}.
  The singularity may be exposed, in the integrable case, by the use of straight field line coordinates, which allow the directional derivative along the magnetic field to be written \mbox{$\sqrt g \, {\bf B}\cdot\nabla \equiv \iotabar \partial_\t + \partial_\phi$}.
  Using a Fourier representation, e.g. \mbox{$\sigma=\sum_{m,n}\sigma_{m,n}\exp(im\t-in\phi)$}, we derive \mbox{$\sigma_{m,n} = -i ( \sqrt g \, \nabla\cdot{\bf j}_\perp )_{m,n} / (m\iotabar-n)$}\mbox{$ + \hat j_{m,n} \, \delta(m\iotabar-n)$}.
  The first term is called the Pfirsch-\Schluter current and has a $1/x$ style singularity at the rational surface, where $x \equiv \iotabar - n/m$.
  The second term, the \mbox{$\delta$}-function current, is generally required to `shield' out resonant magnetic fields that would otherwise destroy the nested family of flux surfaces (a more precise discussion of the $\delta$-function current is provided in \cite{Boozer_Pomphrey_10}).

  In general geometry, the only way to avoid the $1/x$ singular currents is ensure that the pressure gradient is zero in the vicinity of the rational surfaces (or to ensure that no rational surfaces are present).
  As the rational surfaces are dense in space, to avoid the $1/x$ singularities the pressure gradient must be zero everywhere.

  (Despite these concerns near the rational surfaces, VMEC, in particular, does an impressive job of robustly constructing global approximations to 3D equilibria with arbitrary pressure profiles; presumably, this is because VMEC seeks approximations to minima of the {\em global} energy functional, and does not directly seek solutions to \mbox{$\nabla p = {\bf j}\times{\bf B}$} pointwise.)

 \subsection{MRXMHD energy principle} \label{sec:MRXMHD}

  The first step towards constructing the multi-region, relaxed MHD energy functional is to partition space into discrete volumes.
  We introduce a set of nested, toroidal surfaces, \mbox{${\cal I}_{l}$}, for \mbox{$l = 1,2,\ldots N_V$} where \mbox{${\cal I}_{l} \equiv \partial {\cal V}$} for \mbox{$l=N_V$}.
  The energy local to each volume is
 \be \label{eq:localenergyfunctional}
  W_l \equiv \int_{{\cal V}_l} \left( \frac{p}{\gamma-1} + \frac{B^2}{2} \right) dv,
 \ee
 where \mbox{${\cal V}_1$} is the toroid enclosed by \mbox{${\cal I}_1$}, and \mbox{${\cal V}_l$} is the annular volume enclosed by \mbox{${\cal I}_{l-1}$} and \mbox{${\cal I}_{l}$} for \mbox{$l = 2,\ldots N_V$}.

  We again assume that the plasma is in a minimum energy state; however, we allow for the effects of small resistivity: in each \mbox{${\cal V}_l$}, the magnetic field may relax and reconnect (and so topological constraints between the toroidal and poloidal fluxes, the rotational-transform, and the helicity are broken).
  But, in order to retain some control over the equilibria, we consider the \mbox{${\cal I}_l$} to be preserved as {\em ideal barriers} that restrict both pressure transport and field transport.
  Rather than {\em continuously} constraining the topology, the topology is {\em discretely} constrained.
  This, or something equivalent, is required in order to avoid trivial solutions.

  In each \mbox{${\cal V}_l$}, the mass and entropy constraints usually used in ideal MHD do not apply to individual fluid elements, but apply instead to the entire volume, giving the isentropic, ideal-gas constraint,
 \be \label{eq:isentropicpressure}
  p_l V_l^{\gamma} = a_l, 
 \ee
 where \mbox{$V_l$} is the volume of \mbox{${\cal V}_l$} and $a_l$ is a constant.
  The internal energy in ${\cal V}_l$ is \mbox{$\int_{{\cal V}} p_l / (\gamma-1) \, dv$} $=$ \mbox{$a_l V_l^{(1-\gamma)}/(\gamma - 1)$}, and the first variation of this due to a deformation, $\boldxi$, of the boundary is \mbox{$-p \int_{\partial {\cal V}} \boldxi \cdot d{\bf s}$}.

  To constrain the relaxation of the magnetic field in each ${\cal V}_l$, we follow Taylor \cite{Taylor_74}, who argued that the `most conserved' invariant for a weakly resistive plasma is the helicity \cite{Finn_Antonsen_85,Taylor_86},
 \be \label{eq:helicity}
  K_l \equiv \int_{{\cal V}_l} \vec{A} \dotv \vec{B} \, dv,
 \ee
 where $\vec{A}$ is a vector potential, $\vec{B} = \curl\vec{A}$, which we consider to be differentiable and a single-valued function of position.
  The helicity is related to the Gauss linking number: it reflects how `knotted', or `twisted', the magnetic field lines are \cite{Taylor_86,Marsh_96,Berger_99}.
  The helicity in \Eqn{helicity} is not gauge-invariant.
  A gauge-invariant form is constructed by adding the loop integrals $\Delta \psi_p \oint_{\cal S} \mathbf{A}\cdot d\mathbf{l} $ and $\Delta \psi_t \oint_{\cal L} \mathbf{A}\cdot d\mathbf{l}$, where ${\cal S}$ is a poloidal loop on ${\cal I}_{l-1}$ and ${\cal L}$ is a toroidal loop on ${\cal I}_{l}$.

  In each \mbox{${\cal V}_l$}, variations in the pressure and the field, and the geometry of the interfaces, are allowed in order to extremize the energy functional.
  These variations are arbitrary, except for 
   (i) the mass-entropy constraint, \mbox{$p_l V_l^{\gamma} = const$};
  (ii) helicity conservation in each \mbox{${\cal V}_l$};
 (iii) the interfaces must remain tangential to the magnetic field; and
  (iv) the magnetic fluxes are conserved.

  The MRXMHD energy functional is
 \be \label{eq:MRXMHD}
  F = \sum_l \left[ W_l - \frac{\mu_l}{2}\left(K_l-K_{l,o}\right)\right].
 \ee
  The helicity constraint, \mbox{$K_l = K_{l,o}$} where \mbox{$K_{l,o}$} is a given constant, is enforced explicitly by introducing a Lagrange multiplier, \mbox{$\mu_l$}, in each ${\cal V}_l$.
  The flux constraints and the tangentiality condition at the interfaces will be enforced implicitly by constraining the representation of the magnetic field.

  The most general function space for ${\bf B}$ in each volume is space of vector-valued functions whose magnitude is square integrable, i.e. $\vec{B} \in \mathcal{L}^2({\cal V}_l)$, by which it is meant that $B^2 \in \mathcal{L}^1({\cal V}_l)$, and $\mathcal{L}^1({\cal V}_l)$ is the standard notation for the space of integrable scalar functions.
  More precisely, we follow Yoshida et al. \cite{Yoshida_Giga_90,Yoshida_Dewar_12} and restrict $\vec{B}$ to $\mathcal{L}_\sigma^2({\cal V}_l)$, which they define as the subspace of $\mathcal{L}^2({\cal V}_l)$ occupied by divergence-free fields that obey $\vec{B}\cdot {\bf n}=0$ on the boundary.
  Similarly, the pressure is required to be integrable, i.e. $p\in \mathcal{L}^1({\cal V}_l)$.
  While these are the least restrictive spaces required for a weak formulation, in order for the solutions to obey tractable local differential equations almost everywhere we will, after deriving the Euler Lagrange equations for states that extremize \Eqn{MRXMHD}, further restrict the allowed function spaces by assuming that $p$ is piecewise constant and that $\vec{B}$ piecewise satisfies a simple elliptic partial differential equation, which is solved numerically using a mixed Fourier and finite element method.

  The variation in the `local' constrained energy functional, \mbox{$F_l\equiv W_l - \mu_l \left(K_l-K_{l,o}\right)/2$}, due to arbitrary variations in the field, \mbox{$\delta {\bf B} = \nabla \times \delta {\bf A}$}, and arbitrary variations, $\boldxi$, in the interface geometry, is given by
 \be \label{eq:EulerLagrangeEquations}
  \delta F_l &=& \int_{         {\cal V}_l} \left( \nabla \times {\bf B} - \mu_l {\bf B} \right) \cdot \delta {\bf A} \, dv \nonumber \\ 
             &-& \int_{\partial {\cal V}_l} \left( p_l + B^2/2 \right) \boldxi \cdot d{\bf s}.
 \ee

  The variation in the magnetic potential, $\delta\vec{A}$, is free within ${\cal V}_l$, and so within each ${\cal V}_l$ the topology of the field is arbitrary; but at the ${\cal I}_l$ it must obey 
 \be
  \vec{n}\times\delta\vec{A} = -\vec{n}\cdot\boldxi\, \vec{B} + \vec{n}\times\grad\delta g,
 \ee
 so that $\vec{n}\cdot\vec{B} = 0$ remains satisfied; and where $\delta g(\vec{r})$ is the variation in a single-valued gauge potential, $g$, required for generality but physical quantities are invariant with respect to gauge choice.
 The line integrals of $\vec{A}$ along arbitrary loops $\mathcal{L}^{\rm pol}$ and $\mathcal{L}^{\rm tor}$ are related to the poloidal and toroidal magnetic fluxes.

  The enclosed toroidal fluxes in each volume and the poloidal fluxes in each annular region constrain the magnetic field from being trivial.
  We use gauge freedom to specify the loop integrals of $\oint \mathbf{A}\cdot d\mathbf{l}$ on each interface.
  For gauges satisfying these conditions, the difference between the gauge-invariant helicity and the gauge-dependent helicity is a constant.

  The Euler-Lagrange equation for $F$ to be stationary with respect to variations in the magnetic field in each ${\cal V}_l$ is the Beltrami equation, \mbox{$\nabla \times {\bf B} = \mu_l {\bf B}$}.
  The Euler-Lagrange equation for $F$ to be stationary with respect to variations in the interface geometry is that the total pressure must be continuous across the interfaces, \mbox{$[[p+B^2/2]]=0$}.
  States that extremize the MRXMHD energy functional are stepped-pressure equilibria.

  The pressure and tangential field are discontinuous across the interfaces, but these comprise a finite set of measure zero and so $p$ and $B^2$ are both integrable functions: the model is consistent with our goal of constructing weak solutions via an energy-integral approach.
  The discontinuities are easily accommodated for in the numerical discretization; within each volume a continuous, smooth representation for the vector potential is allowed.

  To avoid a problem with `small denominators', as will be discussed below, we will typically enforce the condition that the interfaces have irrational transform.
  The problematic Pfirsch-\Schluter currents are eliminated because the pressure gradient is identically zero across the resonances.
  The $\delta$-function currents are also not present because the topology of field is unrestricted at the rational surfaces, i.e. magnetic islands are allowed to form.

  There are, instead, a finite set of surface currents at the irrational, ideal interfaces, given by $\boldkappa = [[{\bf B}]]\times{\bf n}$, where $[[{\bf B}]]$ is the tangential discontinuity in the field.
  These are required to enforce the topological constraint, the topological constraint in this case being that a noble irrational surface exists; and the topological constraint is required so that the magnetic field matches the given, stepped-pressure profile.
  Given the KAM theorem, this topological constraint is presumably easier to enforce \cite{CBCFLVPFG_04} than forcing a rational flux surface.
  We expect these currents to be dominated by the discontinuity in $p$, and may be thought of as a discrete approximation to the pressure induced currents. 
  These {\em irrational} surface currents may be compared to (but are different from) the $\delta$-function currents shielding at the rational surfaces, which are required in the linearly-perturbed, ideal-equilibrium codes IPEC \cite{Park_Boozer_Glasser_07} and CAS3D \cite{Nuhrenberg_Boozer_03}.
  The $\delta$-function currents at the rational surface currents do not describe the pressure driven $1/x$ singularities.

  We invoke multi-region energy minimization with helicity conservation primarily as a mathematical device to achieve a variational formulation of the restricted equilibrium class, namely stepped-pressure equilibria.
  MRXMHD is, however, a generalization of the variational principle enunciated by Woltjer \cite{Woltjer_58a} to generate linear force-free fields of interest in astrophysics, and developed by Taylor \cite{Taylor_74} to model fusion plasma experiments.

  The success of the Taylor relaxation theory in describing experimental data suggests that the MRXMHD approach may likewise aid physical interpretation of {\em partial} relaxation, reconnection and self-organization in toroidal plasmas supporting a {\em non-trivial} pressure profile.
  Unlike Taylor's globally relaxed model, which gives a constant pressure across the plasma, MRXMHD is only {\em locally} relaxed, i.e. it is {\em partially} constrained; arbitrarily many interfaces may be included, each with an associated ideal $\equiv$ topological constraint.

 \subsection{transform constraint} \label{sec:solvabilitycondition}

  A close examination of the interface force-balance condition, \mbox{$[[p+B^2/2]]=0$}, reveals a Hamiltonian system, which we call the pressure-jump Hamiltonian \cite{MHDvN_10}.
  Let \mbox{$p_-$} and \mbox{${\bf B}_-$} be the pressure and field immediately inside a given interface and \mbox{$p_+$} and \mbox{${\bf B}_+$} be the pressure and field immediately outside.
  By combining (i) the general, covariant representation for the field, \mbox{${\bf B}=B_{s} \nabla s + B_{\t} \nabla \t + B_{\phi} \nabla \phi$}, with (ii) \mbox{$\nabla \times {\bf B} = \mu {\bf B}$}, and (iii) the tangentiality condition, \mbox{${\bf B}\cdot{\bf n}=0$}; we may write \mbox{$B_{\t} = \partial_\t f_\t$} and \mbox{$B_{\phi} = \partial_\phi f$}, and \mbox{$B^2=$}\mbox{$(g_{\phi\phi} f_\t f_\t - 2 g_{\t\phi}f_\t f_\phi + g_{\t\t} f_\phi f_\phi)$}$/$\mbox{$(g_{\t\t}g_{\phi\phi}-g_{\t\phi}g_{\t\phi})$}, where \mbox{$f(\t,\phi)$} is a surface potential and \mbox{$g_{\t\t}$}, \mbox{$g_{\t\phi}$} and \mbox{$g_{\phi\phi}$} are metric elements (local to the interface).
  Now, consider the case where both \mbox{${\bf B}_-$} and the geometry of the interface are known, and we seek a solution for \mbox{${\bf B}_+$} that satisfies \mbox{$H=const$}, where 
 \be \label{eq:pressurejumpHamiltonian}
  H \equiv 2 ( p_- - p_+ ) = B_+^2 - B_-^2.
 \ee
  We may write \mbox{$H \equiv K(\t,\phi,f_\t,f_\phi)+V(\t,\phi)$}, where \mbox{$V \equiv -B_-^2$} is assumed known, and \mbox{$f_\t \equiv \partial_\t f$} and \mbox{$f_\phi \equiv \partial_\phi f$}, where \mbox{$f$} is an as-yet-unknown surface potential for \mbox{${\bf B}_+$}.

  To derive the solvability condition we treat \mbox{$f_\t$} and \mbox{$f_\phi$} as independent quantities (generalized momenta) and recognize \mbox{$H$} as a 2-dof Hamiltonian with a conserved energy, \mbox{$2(p_--p_+)$}.
  Then, $H=const$ along a trajectory given by Hamilton's equations:
 \mbox{$d   \t / dt =     \partial H / \partial f_\t$},
 \mbox{$d f_\t / dt =  -  \partial H / \partial   \t$},
 \mbox{$d   \phi   / dt =     \partial H / \partial f_\phi  $}, and 
 \mbox{$d f_\phi   / dt =  -  \partial H / \partial   \phi  $};
 where \mbox{$t$} is an artificial `time'.
  This system may be reduced to a \mbox{$1\frac{1}{2}$}-dof system by using \mbox{$\phi$} as the time-like integration parameter (always possible if \mbox{$d_t \phi \ne 0$}) and eliminating the integration of \mbox{$f_\phi$} in favor of inverting \mbox{$K(\t,\phi,f_\t,f_\phi)+V(\t,\phi)=2(p_--p_+)$} for $f_\phi$, so that \mbox{$f_\phi$} is assumed to be a function of \mbox{$\t$}, \mbox{$\phi$} and \mbox{$f_\t$}, i.e. \mbox{$f_\phi=f_\phi(\t,\phi,f_\t)$}, where the dependence on \mbox{$2(p_--p_+)$} is implicit.
  The trajectory is then described by \mbox{$\dot \t \equiv d_t \t / d_t \phi $} and \mbox{$\dot f_\t \equiv d_t f_\t / d_t \phi$}, which may be integrated in \mbox{$\phi$} from an initial starting point, \mbox{$( \t, f_\t)$}, on a \Poincare section, e.g. \mbox{$\phi=0$}.
  If the trajectory lies on an invariant surface then it is possible to construct \mbox{$f_\t = f_\t(\t,\phi)$}, and \mbox{$f_\t$} and \mbox{$f_\phi$} recover their interpretation as derivatives of a surface function: there exists a well defined \mbox{$f(\t,\phi)$} such that \mbox{$f_\t=\partial_\t f$} and \mbox{$f_\phi=\partial_\phi f$}.
  That is, if the trajectory lies on an invariant surface, then a solution for \mbox{${\bf B}_+$} that satisfies \mbox{$2 ( p_- - p_+ ) = B_+^2 - B_-^2$} may be constructed.

  An invariant surface can only exist if it avoids the problem of small divisors.
  Note that \mbox{$\dot \t = B^\t / B^\phi $}, so there is a fundamental relationship between the pressure-jump Hamiltonian and the field-line Hamiltonian; and force-balance can only be satisfied if the rotational transform of the interfaces is irrational \cite{MHDvN_10}.

  Many authors have considered sharp boundary equilibria either theoretically or in simplified geometry \cite{BFLMT_86,Kaiser_Salat_94,Lortz_Spies_94,Kaiser_94,Spies_Li_94,Spies_Lortz_Kaiser_01,Spies_03,Kaiser_Uecker_04,Hole_Hudson_Dewar_06,Hole_Hudson_Dewar_07,Hudson_Hole_Dewar_07,Mills_Hole_Dewar_09,HMHD_09}.
  This paper, and our earlier paper \cite{HDHM_11}, represents the first numerical study of toroidal 3D equilibria with multiple Beltrami regions within the plasma.
  The only 3D calculation of which we are aware is the early paper of Betancourt and Garabedian \cite{Betancourt_Garabedian_75}, who consider a free-boundary problem with both the vacuum region and the plasma being Beltrami regions with $\mu = 0$.

  A number of 3D MHD equilibrium codes based on the assumption of continuity and differentiability of $p$ and $\vec{B}$ have been written
  \cite{Betancourt_Garabedian_75,BBG_78,Bauer_Betancourt_Garabedian_82,Betancourt_Garabedian_82,Hirshman_Whitson_83,Bhattacharjee_Wiley_Dewar_84,Bauer_Betancourt_Garabedian_84,Hirshman_Rij_Merkel_86,PMSM_86,Reiman_Greenside_86,HHS_89,Hirshman_Betancourt_91,Taylor_94,SNWNH_06,HSC_11}.
  These have either constrained the magnetic field to be globally integrable;
  have not employed numerical algorithms that explicitly accommodate the singularities in the parallel current at the rational surfaces;
  do not constrain the profiles, and allow the pressure, current and transform profiles to `evolve' during the calculation in a fashion more akin to initial-value, time-evolution codes rather than what is suitable for an equilibrium code;
  have introduced small non-ideal terms, so that the ${\bf B}\cdot\nabla p \ne 0$;
  have ignored the fractal hierarchy of the ergodic invariant sets;
  or employ ill-posed numerical algorithms (e.g. the so-called Spitzer \cite{Spitzer_58} iterative approach \cite{Boozer_84b,Reiman_Greenside_86}, which attempts to invert densely-singular magnetic differential equations \cite{Hudson_10}).
  While they have produced a variety of results, their lack of formal foundations leads them to fall short of the numerical rigor (e.g. demonstration of convergence, quantification of error, estimate of stability \cite{Gardner_Blackwell_92}) available in the axisymmetric case.

  We will now describe the stepped-pressure equilibrium code, SPEC, and demonstrate that the solutions are well defined by presenting convergence studies.

 \section{Numerical Discretization} \label{sec:discretizedenergyfunctional}

  A Fourier representation is employed for all doubly-periodic, scalar functions.
  Even functions, \mbox{$f(-\t,-\z)=f(\t,\z)$}, are written
 \be \label{eq:Fouriersummation}
  f & = & \sum_{n=0}^{N}f_{0,n}\cos(-n N_P \z)\nonumber\\ &+&\sum_{m=1}^{M}\sum_{n=-N}^{N}f_{m,n}\cos(m\t-n N_P \z), 
 \ee
 where \mbox{$N_P$} is the field periodicity.
  The resolution of the Fourier representation is determined by $M$ and $N$, and the total number of Fourier harmonics is \mbox{$N_{MN} \equiv (N + 1) + M ( 2 N + 1)$}.
  The poloidal, \mbox{$\t$}, and toroidal, \mbox{$\z$}, angles are, as yet, arbitrary.
  The Fourier summation will be written concisely as \mbox{$f=\sum_{j}f_j \cos(m_j \t - n_j \z)$}, where \mbox{$(m_1,n_1)=(0,0)$}, etc.
  A similar description is used for odd (i.e. sine) functions, \mbox{$f(-\t,-\z)=-f(\t,\z)$}.

  An initial guess for the geometry of a set of \mbox{$N_V$} nested, toroidal surfaces, ${\cal I}_l$, is assumed given.
  For expedience, we assume stellarator symmetry \cite{Dewar_Hudson_97b}, so that the ${\cal I}_l$ may be described by \mbox{$R(\t,\z) \, \hat R + Z(\t,\z) \, \hat k$}, with
 \be \begin{array}{ccc} R_l(\t,\zeta)&=&\ds \sum_{j}R_{l,j}\cos(m_j \t -n_j \z),\\
                        Z_l(\t,\zeta)&=&\ds \sum_{j}Z_{l,j}\sin(m_j \t -n_j \z), \end{array} \label{eq:stellaratorsymmetricgeometry} 
 \ee
 where \mbox{$\hat R \equiv \cos \phi \,\, \hat i + \sin \phi \,\, \hat j$}, and $\hat i$, $\hat j$ and $\hat k$ are the Cartesian unit vectors.

  To enforce various boundary conditions, it is convenient to use toroidal coordinates, \mbox{$(\s,\t,\z)$}, that are adapted to the interfaces.
  These coordinates are defined inversely via
 \be R=R(\s,\t,\z)\, , \,\, \phi = -\z\, , \,\, Z=Z(\s,\t,\z).\label{eq:coordinates}
 \ee
  The Jacobian of the \mbox{$(s,\t,\zeta)$} coordinates is \mbox{$\sqrt g = R ( R_\s Z_\t - R_\t Z_\s )$}.
  The `lower' metric coefficients, \mbox{$g_{\alpha\beta}$}, are given by \mbox{$g_{\alpha\beta}=R_\alpha R_\beta+Z_\alpha Z_\beta + \delta_{\alpha\beta}R^2$}, where \mbox{$\delta_{\alpha\beta}=1$} if \mbox{$\alpha=\beta=\z$} and \mbox{$\delta_{\alpha\beta}=0$} otherwise.
  The coordinate functions are given by
 \be \begin{array}{ccc} R(\s,\t,\z) & = & \ds \sum_j R_j(\s)\cos(m_j\t-n_j\zeta), \\
                        Z(\s,\t,\z) & = & \ds \sum_j Z_j(\s)\sin(m_j\t-n_j\zeta), \end{array}
 \ee
 where the \mbox{$R_j(\s)$}, \mbox{$Z_j(\s)$} are a piecewise-linear interpolation of the \mbox{$R_{l,j}$} and \mbox{$Z_{l,j}$}.
  (A piecewise-cubic interpolation would give a continuous Jacobian across the interfaces, but this is not required.)
  If the toroidal flux is monotonic increasing, then \mbox{$s\equiv\psi_t$}, normalized to its value at the outermost interface, is a suitable radial coordinate.
  In this case, $s \sim r^2$, where $r$ is a polar-like radial coordinate.
  More generally, we may use the interface label itself as the radial coordinate, i.e. $s_l=l/N_V$.

  In the innermost volume, regularization factors must be included to prevent the interpolated coordinate surfaces from overlapping.
  These factors may be derived by considering an arbitrary, regular (infinitely-differentiable) function, \mbox{$h(x,y)$}, at the origin, $h(x,y) = h + x h_x + y h_y + \frac{1}{2}\left(x^2 h_{xx} + 2 x y h_{xy} + y^2 h_{yy}\right) + \dots$.
  By constructing a Fourier representation, \mbox{$h(r,\t)=\sum_{m}\left[h_m^c(r)\cos(m\t)+h_m^s(r)\sin(m\t)\right]$} where \mbox{$x=r\cos\t$} and \mbox{$y=r\sin\t$}, we obtain after repeated applications of double-angle formulae,
 \be h_{m}(r) = r^m ( a_{0} r^0 + a_{1} r^2 + a_{2} r^4 + a_{3} r^6 +\dots).
 \ee
  So, in the innermost volume, $s\le s_1$, we write \mbox{$R_j(s) = R_{j,1} s^{m_j/2} / s^{m_j/2}_1$}, and similarly for $Z_j(s)$, where $s\sim r^2$.

  In ${\cal V}_l$ that is bounded by \mbox{${\cal I}_{l-1}$} and \mbox{${\cal I}_l$}, a general covariant representation of the magnetic vector potential is 
 \be {\bf \bar A}_l=\bar A_{s,l} \nabla s + \bar A_{\t,l} \nabla \t + \bar A_{\z,l} \nabla \z.
 \ee
  To this add a gauge term, \mbox{$\nabla g_l(s,\t,\z)$}, where \mbox{$g_l$} satisfies
 \be \begin{array}{ccl} \partial_\s g_l(\s ,\t,\z) & = & - \bar A_{\s,l}(\s ,\t,\z) ,\\
 \partial_\t g_l(\s_{l-1},\t,\z) & = & - \bar A_{\t,l}(\s_{l-1},\t,\z)+\psi_{t,l-1},\\
 \partial_\z g_l(\s_{l-1}, 0,\z) & = & - \bar A_{\z,l}(\s_{l-1}, 0,\z)+\psi_{p,l-1},\end{array}
 \ee
 for arbitrary constants, \mbox{$\psi_{t,l-1}$} and \mbox{$\psi_{p,l-1}$}.
  Then \mbox{${\bf A}_l={\bf \bar A}_l+\nabla g_l$} is given by \mbox{${\bf A}_l=A_{\t,l}\nabla\t+A_{\zeta,l}\nabla\z$} with
 \be \begin{array}{ccc} A_{\t,l}(\s_{l-1},\t,\z)&=&\psi_{t,l-1}, \\
                       A_{\zeta,l}(\s_{l-1}, 0,\z) &=&\psi_{p,l-1}. \end{array} \label{eq:gauge}
 \ee

  For stellarator symmetric equilibria, \mbox{$A_{\t,l}$} and \mbox{$A_{\zeta,l}$} may be represented by cosine series,
 \be \begin{array}{ccc} A_{\t,l}(s,\t,\z) & = & \ds \sum_{j}A_{\t,l,j}(s)\cos(m_j\t-n_j\z),\\
                        A_{\zeta ,l}(s,\t,\z) & = & \ds \sum_{j}A_{\zeta ,l,j}(s)\cos(m_j\t-n_j\z), \end{array}
 \ee
 where \mbox{$A_{\t,l,j}(s)$} and \mbox{$A_{\zeta,l,j}(s)$} are represented using finite-elements, as will be described below.

  The toroidal flux is given by
 \be \int_{\cal S} {\bf B} \cdot {\bf ds} &=& \oint_{\partial {\cal S}} {\bf A} \cdot {\bf dl} = 2\pi \; \psi_{t,l-1},
 \ee
 where the surface \mbox{${\cal S}$} is that part of the \mbox{$\z=0$} plane bounded by \mbox{$s=s_{l-1}$}.
  The poloidal flux is given by 
 \be \int_{\cal S} {\bf B} \cdot {\bf ds} &=& \oint_{\partial {\cal S}} {\bf A} \cdot {\bf dl} = 2\pi \; \psi_{p,l-1}.
 \ee
 where the surface \mbox{${\cal S}$} is that part of the \mbox{$\t=0$} plane bounded by \mbox{$s=s_{l-1}$}.

  The boundary condition that the inner interface is a flux surface becomes ${\bf B}\cdot\nabla s=0$, which implies
 \be -m_jA_{\zeta,l,j}(s_{l-1})-n_jA_{\t,l,j}(s_{l-1})=0. \label{eq:innerfluxsurface}
 \ee
  Combing \Eqn{gauge} and \Eqn{innerfluxsurface} we have
 \be A_{\t,l,j}(s_{l-1}) = \left\{ \begin{array}{c@{\quad,\quad}c}\psi_{t,l-1} & j=1, \\ 0 & j>1, \end{array} \right. \label{eq:innergaugeAt}
 \ee
 \be A_{\zeta,l,j}(s_{l-1}) = \left\{ \begin{array}{c@{\quad,\quad}c}\psi_{p,l-1} & j=1, \\ 0 & j>1. \end{array} \right. \label{eq:innergaugeAz}
 \ee
  The condition that the outer interface is a flux surface
 is satisfied by writing 
 \be
 A_{\t,l}(s_{l}) = \partial_\t f_l(\t,\z) \, , \,\,\, A_{\zeta,l}(s_l) = \partial_\z f_l(\t,\z),
 \label{eq:outerboundaryconstraint} 
 \ee
 for arbitrary \mbox{$f$} of the form
 \be f_l=\psi_{t,l} \t + \psi_{p,l} \z + \sum_j f_{l,j}\sin(m_j \t - n_j \z),
 \ee
 and \mbox{$\psi_{t,l}\equiv A_{\t,l,1}(s_l)$} and \mbox{$\psi_{p,l}\equiv A_{\zeta,l,1}(s_l)$}.
  We have
 \be A_{\t,l,j}(s_{l}) = \left\{ \begin{array}{c@{\quad,\quad}c}\psi_{t,l} & j=1, \\ m_j f_{l,j} & j>1, \end{array} \right. \label{eq:outergaugeAt}
 \ee
 \be A_{\zeta,l,j}(s_{l}) = \left\{ \begin{array}{c@{\quad,\quad}c}\psi_{p,l} & j=1, \\ -n_j f_{l,j} & j>1. \end{array} \right. \label{eq:outergaugeAz}
 \ee
 
  The radial dependence of the vector potential harmonics is described using finite-elements.
  A continuous function, \mbox{$f(x)$}, with \mbox{$x\in[0,1]$}, may be approximated using the linear basis functions, \mbox{$\varphi_{L,0}(x)=1-x$} and \mbox{$\varphi_{R,0}(x)=x$}, according to \mbox{$f(x)=f_{L,0}\varphi_{L,0}(x) + f_{R,0}\varphi_{R,0}(x)$}, where \mbox{$f_{L,0}\equiv f(0)$} and \mbox{$f_{R,0}\equiv f(1)$}.
  A piecewise-linear interpolation of the vector potential gives a discontinuous magnetic field in each ${\cal V}_l$.
  While this is legitimate as far as the energy integral is concerned (the magnetic field remains an integrable function), we prefer a smoother interpolation.

  For piecewise-cubic interpolation, the basis functions are \mbox{$\varphi_{L,0}(x) = 2 x^3 - 3 x^2 + 1$} and \mbox{$\varphi_{L,1}(x) = x^3 - 2 x^2 + x$}, and their `reflections', \mbox{$\varphi_{R,p}(x)=(-1)^p \varphi_{L,p}(1-x)$}.
  An arbitrary smooth continuous function is approximated by $f(x)=\sum_{p=0}^{N_D}[f_{L,p}\varphi_{L,p}(x)+f_{R,p}\varphi_{R,p}(x)]$, where \mbox{$f_{L,1}\equiv f^\prime(0)$} and \mbox{$f_{R,1}\equiv f^\prime(1)$} and \mbox{$N_D=1$}.
  For piecewise-quintic interpolation the same expression applies, but with \mbox{$N_D=2$} and 
$\varphi_{L,0}(x) = -6 x^5 + 15 x^4 - 10 x^3 + 1$,
$\varphi_{L,1}(x) = -3 x^5 + 8 x^4 - 6 x^3 + 1$, and 
$\varphi_{L,2}(x) = -\frac{1}{2} x^5 + \frac{3}{2} x^4 - \frac{3}{2} x^3 + \frac{1}{2} x^2$, and their reflections.

  In each ${\cal V}_l$, a regular, radial sub-grid is established,
 \be s_{l,i} = s_{l-1} + i \left( s_i-s_{l-1} \right)/N_l,
 \ee
 for \mbox{$i=0,1,\ldots,N_l$}.
  The resolution, \mbox{$N_l$}, of the radial sub-grid may be different in each ${\cal V}_l$.
  The vector potential harmonics are written
 \be A_{\t,l,j}(s)= \sum_{p=0}^{N_D} \left[ A_{\t,l,j,p,i-1} \; \varphi_{L,p}(x) + A_{\t,l,j,p,i} \; \varphi_{R,p}(x) \right], \nonumber
 \ee
 where \mbox{$x=(s-s_{l,i-1}) / \Delta s_{l}$} with \mbox{$\Delta s_{l}=s_{l,i}-s_{l,i-1}$}.

  The vector potential is completely specified by \mbox{$A_{\t,l,j,p,i}$} and \mbox{$A_{\zeta,l,j,p,i}$}, which are the $p$-th derivatives of the \mbox{$(m_j,n_j)$} harmonics of \mbox{$A_\t$} and \mbox{$A_\z$} on the $i$-th grid-point in the \mbox{$l$-th} annulus.
  Except for the subtlety required to ensure the field is tangential to the outer interface, see \Eqn{outergaugeAt} and \Eqn{outergaugeAz}, these are the independent parameters that describe the vector potential, and thus the magnetic field.

  In the innermost volume, the condition that the field is tangential to the inner interface is replaced (because there is no inner interface) by the condition that the vector potential is analytic at the coordinate origin.
  Assuming $s\sim r^2$, we may enforce regularity at the origin and restrict the gauge by including $s^{m_j/2}$ radial factors with the $A_{\t,l,j,p,i}$ and $A_{\z,l,j,p,i}$, with the boundary conditions
 \be A_{\t,1,j,0,0} & = & 0 \;\;\; \mbox{\rm for all $j$,} \\
     A_{\z,1,j,0,0} & = & 0 \;\;\; \mbox{\rm for $m_j=0$ and $n_j \ne0$.}
 \ee

  The mixed finite-element, Fourier representation of the magnetic vector potential is inserted into the MRXMHD energy integral, $F=\sum_l F_l$, where the `local' energy functional is given by \mbox{$F_l\equiv W_l - \mu_l \left(K_l-K_{l,o}\right)/2$}, where $W_l \equiv \int \left[p/(\gamma-1)+B^2/2\right] dv$ and \mbox{$K_l \equiv \int {\bf A}\cdot{\bf B} \, dv$}.
  With \mbox{${\bf A} = A_\t \nabla \t + A_\z \nabla \z$}, the magnetic field \mbox{${\bf B}=B^\s {\bf e_\s}+B^\t {\bf e_\t}+B^\z {\bf e_\z}$} is 
 \be  \label{eq:contravariantmagneticfield}
 \!\!\!\!\!\!\!\!\!\!\!\! \sqrt g \; {\bf B} & = & (\partial_\t A_\z - \partial_\z A_\t) \; {\bf e_\s} - \partial_\s A_\z \; {\bf e_\t} + \partial_\s A_\t \; {\bf e_\z},
 \ee
 and
 \be B^2 &=& B^\s B^\s \,g_{\s\s} + 2 B^\s B^\t \,g_{\s\t} + 2 B^\s B^\z \,g_{\s\z} \nonumber \\
 &+& B^\t B^\t \,g_{\t\t} + 2 B^\t B^\z \,g_{\t\z} + B^\z B^\z \,g_{\z\z},\nonumber \\
 \sqrt g \; {\bf A} \cdot {\bf B} &=& - A_\t \partial_\s A_\z + A_\z \partial_\s A_\t. \nonumber
 \ee
  After substituting these into the local energy and helicity integrals, the \mbox{$A_{\t,l,j,p,i}$} and \mbox{$A_{\zeta,l,j,p,i}$} are multiplied by terms such as
 \be \label{eq:integralmetrics}
  \intds{l,i-1}{l,i} \intdt \! \intdz \,\, \varphi_{L,p}(s) \,\, \varphi_{R,q}(s) \, f(\s,\t,\z),
 \ee
 where, for example, $f \equiv \sin(m_j\t-n_j\z) \, g_{\alpha\beta} \, \cos(m_k\t-n_k\z)$, and $g_{\alpha\beta}$ are the metric elements (note that the $g_{\alpha\beta}$ depend on the $R_{l,j}$ and $Z_{l,j}$ that define the interface geometry).
  These integrals are computed by constructing a fast Fourier transform of \mbox{$f(\s,\t,\z)$}.
  The integrals over the angles then become trivial and we obtain \mbox{$\bar f(s) \equiv \varphi_{L,p}(s) \varphi_{R,q}(s) \int \! d\t \! \int \! d\z \, f(\s,\t,\z)$}.
  The remaining radial quadrature is approximated using Gaussian integration,
 \be \int_{0}^{1} \!\! ds \, \bar f(s) \approx \sum_{i=1}^{N_G} \omega_i \, \bar f(s_i), 
 \ee
 where the `weights', \mbox{$\omega_i$} and the `abscissae', \mbox{$s_i$}, are chosen to optimize accuracy, and $N_G$ is a numerical resolution parameter that depends on the order of the polynomial being integrated, i.e. $N_G$ depends on the order of the finite-element basis expansion for the \mbox{$A_{\t,l,j,p,i}$} and \mbox{$A_{\zeta,l,j,p,i}$}, and the order of the coordinate interpolation of the $R_{l,j}$ and $Z_{l,j}$.

  In each ${\cal V}_l$, the local energy functional, \mbox{$F_l\equiv W_l - \mu_l \left(K_l-K_{l,o}\right)/2$}, depends on the interface geometry, the vector potential, and various input parameters.
  Specifically, each $F_l$ depends on \mbox{${\bf x} \equiv \{ R_{l,j}, Z_{l,j} \}$}; the enclosed toroidal flux, \mbox{$\Delta \psi_{t,l} \equiv \psi_{t,l}-\psi_{t,l-1}$}; the enclosed poloidal flux, \mbox{$\Delta \psi_{p,l} \equiv \psi_{p,l}-\psi_{p,l-1}$} (except in the innermost volume); the required helicity, \mbox{$K_{l,o}$}; the vector potential, \mbox{${\bf a}_l \equiv \{ A_{\t,l,j,p,i}, A_{\zeta,l,j,p,i} \}$}; and the Lagrange multiplier, \mbox{$\mu_l$}.
  We may indicate the dependence of $F$ as
 \be \label{eq:discretizedFl}
  F_l = F_l[\Delta \psi_{t,l}, \Delta \psi_{p,l},K_{l,o}, {\bf x}; \mu_l, {\bf a}_l ].
 \ee
  In the MRXMHD model, the enclosed fluxes and the helicity in each ${\cal V}_l$ are assumed given, i.e. the $\Delta \psi_{t,l}$, $\Delta \psi_{p,l}$ and $K_{l,o}$ are required input parameters.
  (Note: if the interface rotational-transform constraint is to be given priority over the conservation of poloidal flux and helicity, then $\Delta \psi_{p,l}$ and $K_{l,o}$ must generally be allowed to vary.)
  The computational task is to then find extrema of $F=\sum_l F_l$ with respect to the interface geometry, \mbox{${\bf x} \equiv \{ R_{l,j}, Z_{l,j} \}$}, and the Lagrange multipliers, $\mu_l$, and the vector potentials, \mbox{$\{ A_{\t,l,j,p,i}, A_{\zeta,l,j,p,i} \}$}.

  The basic algorithm is to consider $\mu_l$ and \mbox{${\bf a}_l \equiv \{ A_{\t,l,j,p,i}, A_{\zeta,l,j,p,i} \}$} to be functions of the interface geometry and the $\Delta \psi_{t,l}$, $\Delta \psi_{p,l}$, and $K_{l,o}$.
  That is, first, the Beltrami field, \mbox{$\nabla \times {\bf B} = \mu_l {\bf B}$}, in each ${\cal V}_l$ is constructed.
  We then may write $F_l = F_l[\Delta \psi_{p,l}, K_{l,o}, {\bf x} ]$, where the dependence on $\Delta \psi_{t,l}$ is implicit.
  (Later, for computational expedience, we shall modify this slightly by using $\mu_l$ to parametrize the solutions to the Beltrami fields and remove $K_{l,o}$, so that $F_l = F_l[\Delta \psi_{p,l},\mu_l, {\bf x} ]$.)
  Then, `global' equilibrium states are then constructed by extremizing $F=\sum_l F_l$ with respect to the interface geometry, \mbox{${\bf x} \equiv \{ R_{l,j}, Z_{l,j} \}$}.
 
 \subsection{solving $\nabla\times{\bf B}=\mu {\bf B}$ for ${\bf B}$, given geometry} \label{sec:extremizingdiscretizedstates}

  Assuming that the geometry of the interfaces, \mbox{${\bf x}$}, is given, there are various numerical methods that may be employed to construct the extremizing fields.
  The first method is the standard Lagrange multiplier approach: a multi-dimensional Newton method is used to find an {\em extremum} of the local constrained energy functional.
  The solution satisfies
 \be \frac{ \partial F_l }{ \partial {\bf a}_l }  =  0 \; , \;\; 
     \frac{ \partial F_l }{ \partial  \mu_l  }  =  0 \;,
 \ee
 where, in addition to \mbox{${\bf a}_l \equiv \{ A_{\t,l,j,p,i}, A_{\zeta,l,j,p,i} \}$}, the Lagrange multiplier is explicitly treated as an independent degree of freedom and must be adjusted to satisfy the helicity constraint.
  
  This approach cannot distinguish states that \emph{minimize} $W_l$ from states which are saddle points or local maxima of $W_l$.
  When bifurcated solutions exist, i.e. when there exist multiple stationary points of $W_l$ and hence $F_l$, a gradient-descent algorithm such as sequential quadratic programming \cite{Nocedal_Wright_99} may instead be used to ensure that the constructed solution is strictly a local minimum of $W_l$ subject to the constraint $K_l = K_{l,o}$

  Another method for constructing the Beltrami fields, on which we hereafter concentrate, is to assume each ${\bf B}_l$ is parametrized by the enclosed fluxes and $\mu_l$.
  The required value for the helicity, $K_{l,o}$, may be dropped from the local energy functional, to obtain \mbox{$F_l \equiv W_l - \mu_l K_l / 2 $}, and we write
 \be
  F_l = F_l[\Delta \psi_{p,l}, \mu_l, {\bf x} ; {\bf a}_l ],
 \ee
 where the dependence on $\Delta \psi_{t,l}$ is implicit.

  The local energy functional, $F_l$, is quadratic in the \mbox{$A_{\t,l,j,p,i}$} and \mbox{$A_{\zeta,l,j,p,i}$}, and the `local' equilibrium condition, \mbox{$\partial F_l / \partial {\bf a}_l = 0$}, gives a system of linear equations to be solved for the vector potential.
  We call this the Beltrami linear system, as it is analogous to $\nabla \times {\bf B} = \mu_l {\bf B}$ and can be represented as
 \be \label{eq:BeltramiLinearSystem}
  {\bf G} \cdot {\bf a} = {\bf c},
 \ee
 where the matrix ${\bf G}$ depends on the geometry, the fluxes and $\mu_l$, i.e. ${\bf G} = {\bf G}[\Delta \psi_{p,l}, \mu_l, {\bf x} ]$; and similarly for the right-hand-side vector, ${\bf c}$.
  This system is inhomogeneous (i.e. ${\bf c}$ is non-trivial) because of the `driving' terms \mbox{$A_{\t,1,l,0,0}=\psi_{t,l}$} and \mbox{$A_{\zeta,1,l,0,0}=\psi_{p,l}$}.
  Given ${\bf x}$, there is a two-dimensional family of solutions; each solution is parametrized by $\Delta \psi_{p,l}$ and $\mu_l$.

  There is an abundance of numerical methods and `canned' numerical routines available for solving linear equations, and any mathematical structure present can be exploited.
  For example, usually the matrix ${\bf G}$ is positive definite and it is typically very sparse, and an initial `guess' for the solution is often available (particularly so during an iterative calculation).
  Employing numerical methods that exploit the sparsity and positive definite-ness can significantly improve code performance.

  It will be efficient to know how the Beltrami field in a given volume varies with small variations in both the input parameters and the interface geometry.
  In equilibria that are globally constrained by ideal MHD, variations in the magnetic field are related to variations in geometry via $\delta {\bf B} = \nabla \times ( \boldxi \times {\bf B})$.
  In our case, we can compute the change in the vector potential resulting from an infinitesimal change in $\Delta \psi_{p,l}$, or in $\mu_l$, or in the interface geometry, ${\bf x}$, using matrix perturbation theory, $({\bf G} + d{\bf G})\cdot({\bf a} + d{\bf a}) = ({\bf c} + d{\bf c})$, so that to lowest order
 \be \label{eq:matrixperturbation}
  {\bf G} \cdot d{\bf a} = d{\bf c} - d{\bf G} \cdot {\bf a}.
 \ee
  The infinitesimal variations, $d{\bf G}$ and $d{\bf c}$, resulting from infinitesimal variations in $\Delta \psi_{l,p}$ and $\mu_l$ are rather simple to write down because $\Delta \psi_{l,p}$ and $\mu_l$ just appear as factors multiplying various geometric quantities in ${\bf G}$ and ${\bf c}$.
  The derivatives with respect to the $R_{l,j}$ and $Z_{l,j}$ are more complicated as these involve differentiating \Eqn{integralmetrics}. 

  The Beltrami field in ${\cal V}_l$ depends on the geometry of both the `inner' interface, i.e. the $R_{l-1,j}$ and $Z_{l-1,j}$, and the `outer' interface, i.e. the $R_{l,j}$ and $Z_{l,j}$; giving a total of $4 N_{MN} - 2$ geometrical degrees of freedom, where $N_{MN}$ describes the Fourier resolution.
  In preference over repeatedly inverting the Beltrami matrix $4 N_{MN} - 2$ times, which would be the case if finite-differences for example were used to compute the change in the Beltrami fields, we instead first compute a Cholesky factorization of ${\bf G}$, i.e. ${\bf G}={\bf L}\, {\bf L}^T$.
  Then, the solution to \Eqn{matrixperturbation} is efficiently given by ${\bf L}\cdot{\bf y}={\bf b}$, where ${\bf b}\equiv d{\bf c} - d{\bf G} \cdot {\bf a}$, and ${\bf L}^T \cdot d{\bf a}={\bf y}$. 

  The helicity, $K_l \equiv \int {\bf A}\cdot{\bf B} \;dv$, depends on the solution to \Eqn{BeltramiLinearSystem}, which in turn depends on $\mu_l$.
  The helicity constraint, $K_l = K_{l,o}$, can be enforced by suitably adjusting $\mu_l$.
  (This is not always possible; this is only for configurations in which $\mu_l$ parametrizes states with different helicity.)

 \subsection{transform constraint, noble irrationals} \label{sec:nobletransform}

  The rotational-transform constraint can similarly be enforced.
  If only the Beltrami field within a single annulus is to be constructed, then there is no constraint on the allowed values of the transform on the interfaces.
  Recall however, that if the Beltrami fields in multiple volumes are to be consistently nested together in a fashion that satisfies force balance, an analysis of the pressure-jump Hamiltonian derived from $[[p+B^2/2]]=0$ shows that the interfaces should have irrational transform.

  We restrict attention to noble irrationals \cite{Niven_56}, which play an important role \cite{Meiss_92} in the theory of chaos as invariant KAM surfaces with noble transform are most likely to survive chaos-inducing perturbations \cite{Greene_79}.
  A noble irrational is obtained as the limit of an infinite, alternating path down a Farey tree, which is constructed as follows.
  Begin with a pair of rationals, \mbox{$p_1/q_1$} and \mbox{$p_2/q_2$}, which should be neighboring, i.e. \mbox{$|p_1 q_2 - p_2 q_1| = 1$}, and without loss of generality we assume that $p_1/q_1<p_2/q_2$.
  A Farey tree is constructed by successively constructing the mediants, defined as \mbox{$p/q=(p_1+p_2)/(q_1+q_2)$}.
  This is guaranteed to lie between the original `parent' rationals and so splits the original interval into left, \mbox{$[p_1/q_1,p/q]$}, and right, \mbox{$[p/q,p_2/q_2]$}, sub-intervals.
  The mediant is neighboring to both parents, and the construction of the Farey tree proceeds iteratively.
  An infinite, alternating path down the Farey tree is a sequence of mediants for alternately the left and right subintervals.
  Sequences of this type converge to noble irrationals, which have continued fraction representations that terminate in an infinite sequence of $1$'s \cite{Meiss_92}.
  It is easy to see that alternating paths give Fibonacci sequences for the numerator and denominator of the successive rationals.
  For example, beginning from \mbox{$p_1/q_1=0/1$} and \mbox{$p_2/q_2=1/1$} and constructing an alternating path of mediants we obtain the sequence $\frac{0}{1}$, $\frac{1}{1}$, $\frac{1}{2}$, $\frac{2}{3}$, $\frac{3}{5}$, $\frac{5}{8}$, $\dots$
  This allows noble irrationals to be written in the concise form $\iotabar(p_1,q_1,p_2,q_2) = (p_1+\gamma \,p_2)/(q_1+\gamma \,q_2 )$, where the golden-mean, \mbox{$\gamma = ( 1 + \sqrt{5} ) / 2$}, is the limiting ratio of successive terms, \mbox{$\gamma=F_{n+1}/F_{n}$} as \mbox{$n\rightarrow\infty$}, of the Fibonacci sequence beginning from \mbox{$F_0=0$} and \mbox{$F_0=1$}.

  The poloidal angle parametrization of the interfaces is, at this stage, arbitrary; it is not required, and will not be required, that the field lines are `straight'.
  We may, however, construct the straight-field-line angle on the interfaces, given the field, by calculating the angle transformation, \mbox{$\t_s \equiv\t+\lambda(\t,\z)$}, such that
 \be \label{eq:straightfieldlineangle}
  \frac{{\bf B}\cdot\nabla\t_s}{{\bf B}\cdot\nabla\z}=\iotabar,
 \ee
 where \mbox{$\iotabar$} is, as yet, an unknown constant to be determined.
  We restrict attention to angle transformations of the form \mbox{$\lambda=\sum_j \lambda_j \sin(m_j \t - n_j \z)$}, which preserves stellarator symmetry but is otherwise general.
  The Fourier resolution of the angle transformation is independent of the Fourier resolution, $M$ and $N$, used to represent the interfaces and Beltrami fields, and typically we use an enhanced Fourier resolution for $\lambda$.
  With \Eqn{contravariantmagneticfield} and using \mbox{$B^s=0$}, \Eqn{straightfieldlineangle} becomes $A_\t^\prime \, \partial_\z \lambda - A_\z^\prime \, \partial_\t \lambda - A_\t^\prime \, \iotabar = A_\z^\prime$, where the prime indicates radial derivative.
  By equating coefficients, we obtain a system of linear equations for the unknowns, \mbox{${\boldlambda}=(\iotabar,\lambda_2,\lambda_3,\dots )^T$}, which is represented as a matrix equation,
 \be \label{eq:TransformLinearSystem}
  {\bf \Lambda} \cdot \boldlambda = {\bf d}, 
 \ee
 where \mbox{${\bf \Lambda}$} and \mbox{${\bf d}$} depend on the \mbox{$A_\t^\prime$} and \mbox{$A_\z^\prime$} harmonics at the interfaces.
  Solving this linear-system determines $\boldlambda$, which gives the rotational-transform on the interface, namely \mbox{$\iotabar$}.

 \insertdblfigure{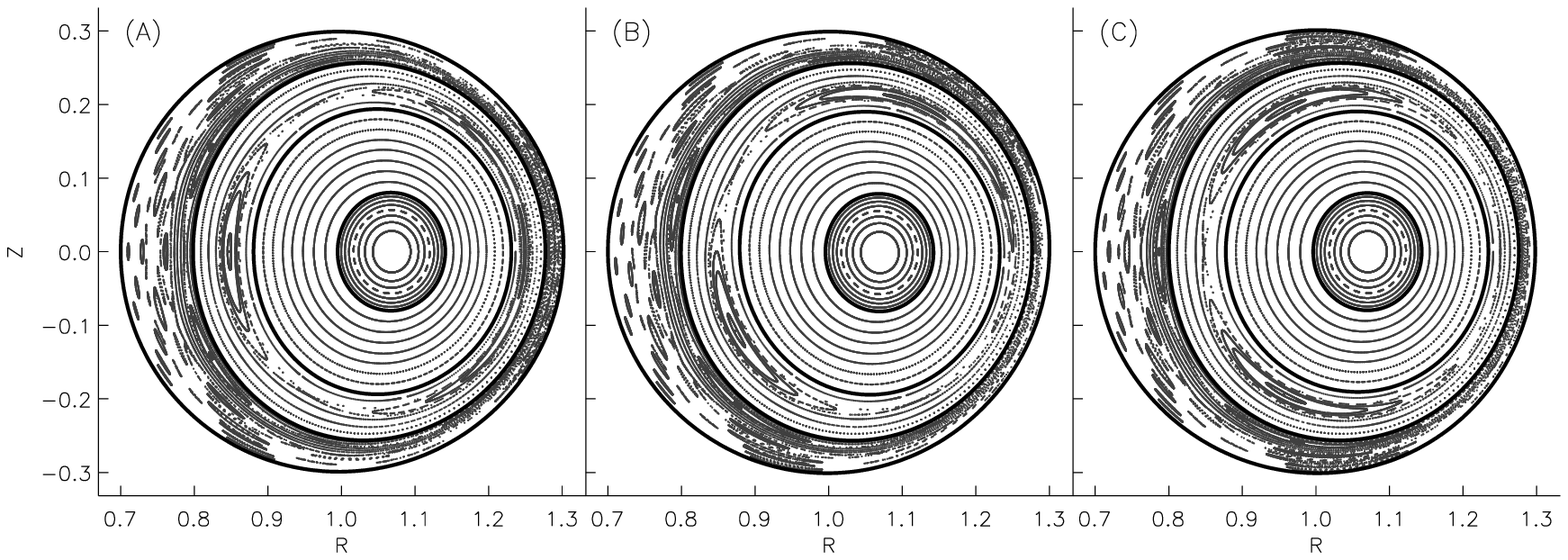}{\Poincare plot showing the Beltrami fields in multiple, nested volumes and the ideal interfaces (thick lines) for the perturbed equilibrium on the cross sections (A) $\z=0$; (B) $\z=\pi/2$; and (C) $\z=\pi$.}{BeltramiPoincare}

  Considering the geometry of the interfaces and the enclosed toroidal flux in each ${\cal V}_l$ to be fixed, each Beltrami field depends only on \mbox{$\mu_l$} and \mbox{$\Delta \psi_{p,l}$}.
  We thus have two degrees-of-freedom, and we must satisfy two constraints; these constraints being that the field in ${\cal V}_l$ provides the required rotational transform on the inner interface, ${\cal I}_{l-1}$, and on the outer interface, ${\cal I}_{l}$.
  (In the innermost volume, ${\cal V}_1$, there is only one degree of freedom, namely $\mu_1$, and only one constraint, namely that the field provides the required rotational transform on ${\cal I}_1$.)
  Defining the function \mbox{$f(\mu_l, \Delta \psi_{p,l})=(\iotabar_{inn}-\iotabar_{l-1}, \iotabar_{out}-\iotabar_{l} )$}, where \mbox{$\iotabar_{inn}$} and \mbox{$\iotabar_{out}$} are the transforms determined from solving \Eqn{TransformLinearSystem} on the inner and outer interface for the magnetic field parametrized by \mbox{$(\mu_l, \Delta \psi_{p,l})$}, and \mbox{$\iotabar_{l-1}$} and \mbox{$\iotabar_{l}$} are prescribed input values, we employ a simple Newton method to set $f(\mu_l, \Delta \psi_{p,l})=0$.
  Typically, if a reasonable guess is provided, this converges in one or two iterations.
  Matrix perturbation methods are used to compute the derivatives: the infinitesimal variation, $d\boldlambda$, resulting from an infinitesimal variation in $\Delta \psi_l$ or $\mu_l$ is given by ${\bf \Lambda} \cdot d{\boldlambda} = d{\bf d} - d{\bf \Lambda} \cdot \boldlambda$, and $d{\bf \Lambda}$ depends on ${\bf \Lambda}$ and $d{\bf G}$ via the chain rule; and similarly for $d{\bf d}$.
  This search is computationally efficient: the integral metric elements, \Eqn{integralmetrics}, do not need to be recomputed if the geometry does not change; and the matrix ${\bf \Lambda}$ is very sparse, and sparse linear algorithms are employed.

  In the MRXMHD model, the poloidal flux and helicity are assumed given.
  In the stepped-pressure model, the interface rotational transforms are constrained.
  (In both cases, the toroidal flux is constrained.)
  In either case, given the interface geometry, the Beltrami field in each volume that satisfies these constraints is unique -- except for the possibility of bifurcations, which we do not consider in this article.
  So, with this implicit, the dependence of each $F_l$ on the degrees of freedom is reduced to $F_l = F_l[{\bf x}]$.

  The task of constructing global equilibrium solutions is now the standard mathematical problem of finding extrema of the global energy functional, $F=F[{\bf x}]$.
  Before describing the two basic approaches we have adopted for this, namely a preconditioned conjugate gradient algorithm for minimizing the global energy functional and a Newton-style algorithm for finding a zero of the multi-dimensional force-balance vector, we first present a convergence study illustrating that the Beltrami field in each ${\cal V}_l$ may be constructed to arbitrary accuracy.

 \subsection{illustration of Beltrami fields} \label{sec:illustrationBeltrami}

  For illustration, we show the Beltrami fields consistent with a multi-region equilibrium.
  The equilibrium is defined by the toroidal flux enclosed by each interface, the pressure in each ${\cal V}_l$, and the interface transforms, and these are all given in \Tab{toroidalfluxandtransformconstraints}.
  The outer boundary is given by \mbox{$R=1 + r \cos(\t)$} and \mbox{$Z=r \sin(\t)$}, with $r=0.3 + \delta \cos(2\t-\z) + \delta \cos(3\t-\z)$ and $\delta=10^{-3}$.
  This choice of perturbation induces `primary' islands at the $\iotabar=1/2$ and $\iotabar=1/3$ rational surfaces.
  The interior boundaries are consistent with force balance.
  The interface cross sections and \Poincare plots of the Beltrami fields are shown in \Fig{BeltramiPoincare}.
  Because of toroidal and poloidal coupling, magnetic islands (and irregular field lines) will form at all rational surfaces within the rotational transform range.

 \begin{table}
 \caption{Flux and transform constraints}
 \begin{center}
 \begin{tabular}{ccccc} \hline
 $ l $ & $ \psi_{t,l} $ & $p_l$ & $ \iotabar_{l}$                             &                     \\ \hline
 $  1 $ & $ 0.05950 $ & $ 0.94168 $ & $(  5 + \gamma \,  6 )/(  6 + \gamma \,  7 ) $ & $ =   0.848898\dots $ \\
 $  2 $ & $ 0.35098 $ & $ 0.63872 $ & $(  1 + \gamma \,  2 )/(  2 + \gamma \,  3 ) $ & $ =   0.618034\dots $ \\
 $  3 $ & $ 0.64902 $ & $ 0.25740 $ & $(  1 + \gamma \,  1 )/(  2 + \gamma \,  3 ) $ & $ =   0.381966\dots $ \\
 $  4 $ & $ 1.00000 $ & $ 0.04106 $ & $(  1 + \gamma \,  1 )/(  9 + \gamma \, 10 ) $ & $ =   0.103971\dots $ \\
 \end{tabular}
 \label{tab:toroidalfluxandtransformconstraints}
 \end{center}
 \end{table}

  \insertdblfigure{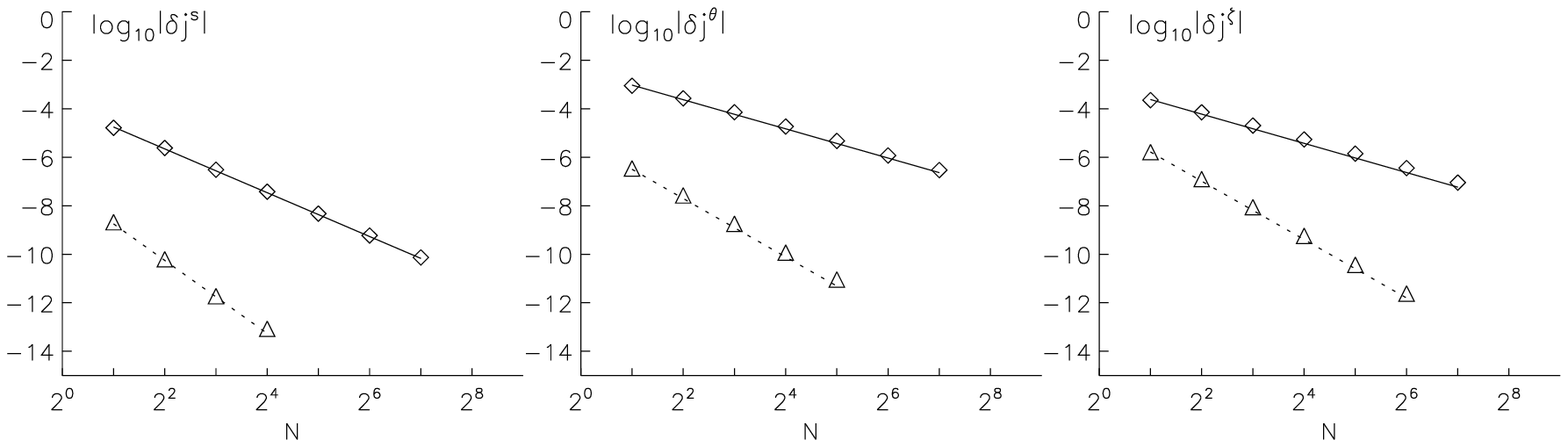}{Scaling of components of error, $\delta {\bf j}\equiv{\bf j}-\mu{\bf B}$, with respect to radial resolution. The diamonds are for the $n=3$ (cubic) basis functions, the triangles are for the $n=5$ (quintic) basis functions. The solid lines have gradient -3,-2 and -2, and the dotted lines have gradient -5,-4 and -4.}{curlerror}

  With finite resolution, the equation \mbox{$\nabla \times {\bf B}=\mu {\bf B}$} can of course only approximately be solved.
  However, given a smooth boundary, the solution to the Beltrami equation is well posed, and so the numerical error can be made arbitrarily small.
  Assuming that the Fourier resolution is sufficient to ensure the numerical error results from the finite-ness of the radial discretization, an \mbox{$n$}-th order approximation to the vector potential yields an error \mbox{${\cal O}(h^{n+1})$}, where \mbox{$h$} is the radial sub-grid size.
  The Fourier harmonics of the contravariant components of ${\bf B}$ are
 \mbox{$\left(\sqrt g B^\s\right)_{m,n}=-m A_{\t,m,n} -n A_{\z,m,n} $},
 \mbox{$\left(\sqrt g B^\t\right)_{m,n}= A_{\z,m,n}^\prime $}, and
 \mbox{$\left(\sqrt g B^\z\right)_{m,n}= - A_{\t,m,n}^\prime$},
 where the prime denotes radial derivative.
  Radial derivatives reduce the order of the error, and ${\bf B}=\nabla \times {\bf A}$ is generally an order less accurate that ${\bf A}$ itself; with the exception of $B^\s$, which remains accurate to \mbox{${\cal O}(h^{n+1})$} as no radial derivatives are involved.
  Before computing \mbox{$\nabla \times {\bf B}$}, the `contravariant' components must be `lowered', \mbox{$B_\alpha=\sum_\beta g_{\alpha\beta}B^\beta$}, and the error in \mbox{$B_\s$}, \mbox{$B_\t$}, \mbox{$B_\z$} are each \mbox{${\cal O}(h^{n})$}.
  The Fourier harmonics of the contravariant components of ${\bf j}\equiv\nabla\times{\bf B}$ are computed similarly,
 and the error in $j^\s$, $j^\t$ and $j^\z$ are \mbox{${\cal O}(h^{n})$}, \mbox{${\cal O}(h^{n-1})$} and \mbox{${\cal O}(h^{n-1})$}, respectively.
  The components of the error, \mbox{${\bf j}-\mu {\bf B}$}, are quantified by
 \be
  |\delta j^\alpha|&=&\left[ \sum_{m,n} \left[ (\sqrt g j^\alpha)_{m,n} - \mu (\sqrt g B^\alpha)_{m,n} \right]^2/N_{MN}\right]^{1/2}, \nonumber
 \ee
 for $\alpha=\s,\t$, and $\zeta$.
  These quantities are shown as a function of radial sub-grid resolution in \Fig{curlerror}, for the field in ${\cal V}_3$ of the equilibrium shown in \Fig{BeltramiPoincare}.
  The expected error scalings, \mbox{$|\delta j^\s|\sim {\cal O}(h^{n})$}, \mbox{$|\delta j^\t|\sim{\cal O}(h^{n-1})$} and \mbox{$|\delta j^\z|\sim{\cal O}(h^{n-1})$}, are confirmed for both the cubic, \mbox{$n=3$}, and quintic, \mbox{$n=5$}, finite-element representations.

  Note that at no point does the algorithm depend of resolving the fractal structure of phase space.
  The vector potential in each ${\cal V}_l$ is a smooth function, both as a function of position within a given volume, and as a function of interface geometry.

  Before proceeding to the task of piecing together multiple, nested Beltrami fields to obtain global, non-trivial equilibria, we must tie down a `loose-end' regarding the Fourier representation of the interfaces.

 \subsection{spectral condensation} \label{sec:spectralcondensation}

  To construct global equilibria, the $R_{l,j}$ and $Z_{l,j}$ describing the interface geometry will be varied to extremize the energy functional and/or satisfy force balance.
  Tangential geometric variations merely change the angular parametrization of the interfaces and do not change the interface geometry, and so do not affect the energy functional, but {\em do} alter the $\{ R_{l,j}, Z_{l,j} \}$.
  This freedom may be exploited to obtain a {\em preferred} angle parametrization.

  A numerically insightful choice \cite{Hirshman_Meier_85,Hirshman_Breslau_98} is to choose the angle that minimizes the `spectral width', and so obtain the most accurate representation of the interface geometry for a given Fourier resolution.
  We define the spectral width as
 \be \frac{1}{2}\left.\sum_j\right. (m_j^p+n_j^q) \left(R_j^2 + Z_j^2\right), \label{eq:spectralwidth}
 \ee
 where \mbox{$p$} and $q$ are arbitrary integers required as input.

  The toroidal angle has already been constrained, \mbox{$\z\equiv-\phi$}, and the geometry of the interfaces is to be constrained by force balance, so we are left to minimize the spectral width with respect to {\em poloidal} variations, \mbox{$\delta R = \partial_\t R \, \delta u$} and \mbox{$\delta Z = \partial_\t Z \, \delta u$}.
  To preserve stellarator symmetry we restrict attention to odd functions, \mbox{$\delta u=\sum_j u_j \sin(m_j\t-n_j\z)$}.
  The variations in the Fourier harmonics, $R_j$ and $Z_j$, are given by \mbox{$\delta R_j = \oint \oint R_\t \delta u \, \cos(m_j\t-n_j\z) d\t d\z$} and \mbox{$\delta Z_j = \oint \oint Z_\t \delta u \sin(m_j\t-n_j\z)d\t d\z$}, where \mbox{$R_\t \equiv \partial_\t R$} and \mbox{$Z_\t \equiv \partial_\t Z$}.
  The first variation in the spectral width is
 \be \label{eq:spectralcondensationEL}
  \oint \!\!\! \oint d\t d\z\left(R_\t X + Z_\t Y\right) \delta u,
 \ee
 where \mbox{$X = \sum_j (m_j^p+n_j^q) R_j \cos(m_j\t-n_j\z)$} and \mbox{$Y = \sum_j (m_j^p+n_j^q) Z_j \sin(m_j\t-n_j\z)$}.
  The spectral width is decreased along $\delta u = - I$, where \mbox{$I \equiv R_\t X + Z_\t Y$}, and is extremized when \mbox{$I = 0 $}.

 \subsection{illustrations of global equilibria} \label{sec:benchmarks}

  We have implemented two numerical methods for finding global equilibria.
  The first is a minimization algorithm: we seek minima of \mbox{$F[{\bf x}]\equiv \sum_l \int_{\cal V} [p_l/(\gamma-1)+B^2/2] dv$} using a preconditioned, conjugate gradient algorithm, where in each volume $p_l V^\gamma = a_l$ is constant, and the field satisfies $\nabla \times {\bf B} = \mu_l {\bf B}$.
  The gradient of \mbox{$F$} with respect to the \mbox{$R_{l,j}$} and \mbox{$Z_{l,j}$}, with the fluxes and helicity constrained, can be derived from \Eqn{EulerLagrangeEquations} by recalling that the displacement is $\boldxi \equiv \delta R \, {\hat R} \, + \delta Z \, {\hat k}$ and the surface element is $d{\bf s} \equiv - R Z_\theta \, \hat R + (Z_\theta R_\zeta - R_\theta Z_\zeta) \, \hat \phi + R R_\theta \, \hat z$.
  The derivatives of $F$ are
 \be
  \! \frac{\partial F}{\partial R_{l,j}} \! & \! = \! & \! - \left( [[ p + B^2/2 ]]_l \, R \, Z_\t \right)_j - (R_\t I_l )_j, \\
  \! \frac{\partial F}{\partial Z_{l,j}} \! & \! = \! & \! + \left( [[ p + B^2/2 ]]_l \, R \, R_\t \right)_j - (Z_\t I_l )_j,
 \ee
 where the $(R_\t I_l)_j$ and $(Z_\t I_l)_j$ terms are included to reduce the spectral width.

  The second approach for finding global equilibria is a globally-convergent, multi-dimensional Newton method.
  We seek a zero of the `force-balance' vector, constructed as follows.
  Global force balance is satisfied when the total pressure discontinuity across each interface is zero, $[[p+B^2/2]]=0$. 
  The Fourier representation of the interfaces minimizes the spectral width when \mbox{$I$} is zero.
  Thus, we construct a `constraint' vector, ${\bf f}({\bf x})$, by collecting together the harmonics $[[p+B^2/2]]_{l,j}$ and $I_{l,j}$. 

  \insertsglfigure{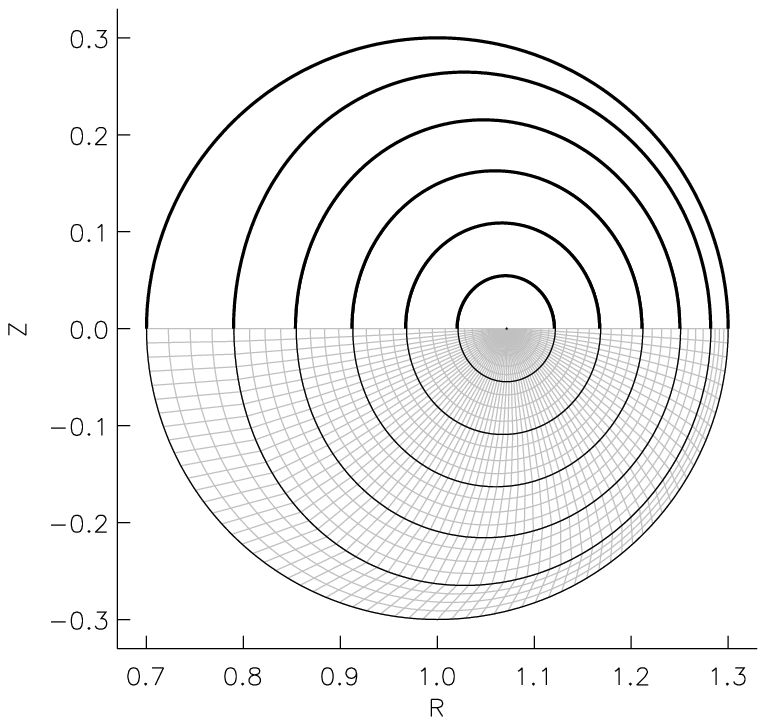}{Comparison between the SPEC interfaces, with \mbox{$N_V=6$}, and the corresponding VMEC surfaces (thick lines, upper half); and the SPEC radial sub-grid (lower half).}{approximationtosmoothPoincare}

  \insertsglfigure{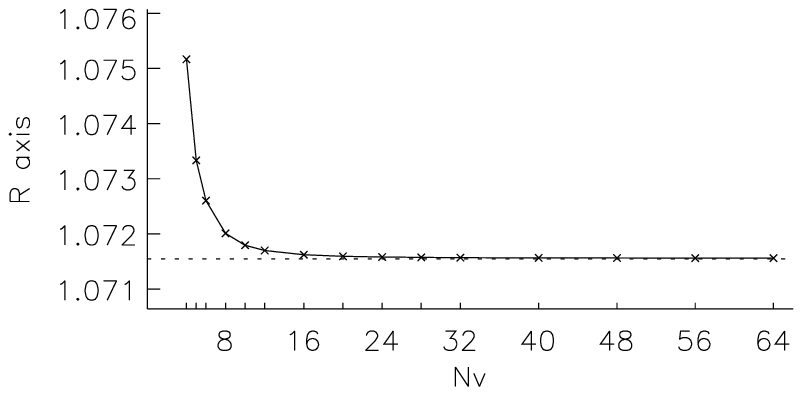}{Location of magnetic axis as computed by SPEC against resolution, $N_V$, of stepped-pressure approximation to smooth pressure profile. The dotted line is the location of magnetic axis for a high resolution VMEC calculation.}{axisconvergence}

  \insertsglfigure{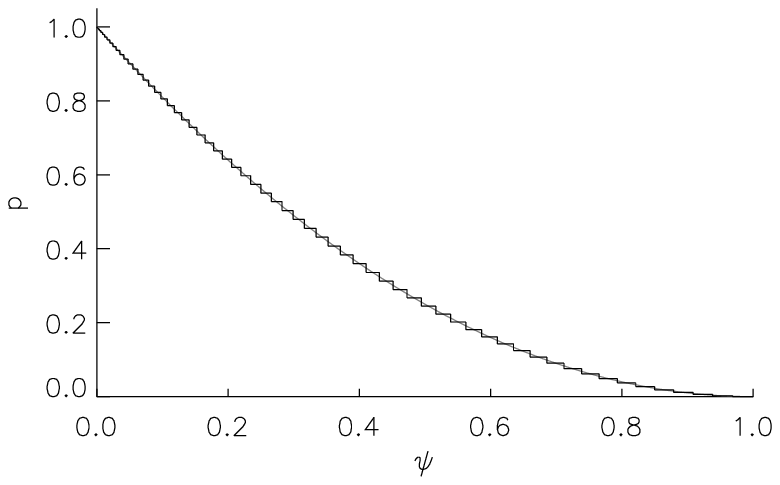}{Stepped pressure approximation, with $N_V=64$, to smooth pressure profile.}{steppedsmooth}

  Within the stellarator symmetric representation we have employed, $R$ is an even function of $(\t,\z)$, and $Z$ is odd.
  Force-balance, $[[p+B^2/2]]$, and the spectral minimization condition, $I$, are similarly even and odd functions.
  So, after suitably truncating $[[p+B^2/2]]$ and $I$ to match the truncated Fourier representation of the interface geometry, the number of constraints equals the number of degrees-of-freedom.

  Expanding about an arbitrary point, \mbox{${\bf f}({\bf x}+\delta {\bf x}) \approx {\bf f}({\bf x}) + \nabla {\bf f}({\bf x}) \cdot \delta {\bf x}$}, the Newton correction required to find the equilibrium point, ${\bf f}=0$, is given by $\delta {\bf x} = -\nabla {\bf f}^{-1} \cdot {\bf f}.$
  The Jacobian matrix, $\nabla {\bf f}$, describes how the Beltrami fields change when the geometry is changed (more precisely, how $B^2$ on the interfaces changes when the ${\cal I}_l$ change) and is computed in parallel using matrix perturbation methods as described above.

  In the following, we use this method for constructing global solutions.
  In all calculations presented, either explicitly or implicitly, for any given Fourier and radial sub-grid resolution, the error in force balance, $|f|$, and the error in `position', $|\delta x|$, are less than $10^{-12}$.
 
  Before presenting illustrations of non-axisymmetric global equilibria, we first present a comparison of stepped-pressure equilibria to axisymmetric MHD equilibria with smooth profiles, the latter computed by VMEC \cite{Hirshman_Rij_Merkel_86}.
  SPEC is intended for computing equilibria with partially chaotic fields and only admits stepped-pressure profiles.
  In contrast, VMEC globally constrains the field to be integrable, and so admits (and assumes) smooth profiles.
  As it is the profiles that define an equilibrium, VMEC and SPEC will differ.
  To obtain agreement, it is required to perform a convergence study in the pressure profile: as the number of steps in the stepped profile increases, the stepped profile will better approximate a smooth profile.
 
  We consider an equilibrium with boundary described by \mbox{$R(\t,\z)  =  1.0  +  0.3 \cos \t$} and \mbox{$Z(\t,\z)  =          0.3 \sin \t$}.
  For the (smooth) pressure profile we take \mbox{$p(\psi) = p_0(1 - 2 \psi + \psi^2)$}, where $\psi$ is the normalized toroidal flux and $p_0$ is a scaling factor chosen to give a Shafranov shift about one-third the minor radius.
  For the (smooth) transform profile we take \mbox{$\iotabar=\iotabar_0-\iotabar_1 \psi$}, where \mbox{$\iotabar_0 = $} \mbox{$(8+9\gamma)/$} \mbox{$(9+10\gamma)$} and \mbox{$\iotabar_1 = $}   \mbox{$\iotabar_0-$}   \mbox{$(1+1\gamma)/$}   \mbox{$(9+10\gamma)$}.
  We use high Fourier resolution, $M=16$, and high radial sub-grid resolution; this will ensure that any discrepancy results from finite $N_V$, and so the error will decrease as $N_V$ is increased.

  As input to SPEC, we must `discretize' the pressure and transform profiles.
  Earlier we recalled that invariant surfaces of Hamiltonian systems with noble rotational transform are the most likely to survive chaos-inducing perturbations, and so the interface transforms should be strongly irrational; however, in the axisymmetric and therefore integrable case, the interface transforms may be chosen arbitrarily.
  For convergence studies, it is preferable to have interfaces regularly spaced in radius and we choose the interface rotational transforms $\iotabar_l = \iotabar(\psi_l)$, where $\sqrt \psi_l = l/ N_V$.
  We discretize the pressure profile in such a way as to piecewise conserve the integrated pressure,
 \be 
  p_l \int_{\psi_{l-1}}^{\psi_l} \!\!\!\!\!\!\! d\psi = \int_{\psi_{l-1}}^{\psi_l} \!\!\!\!\!\! p(\psi) \, d\psi.
 \ee

  For illustration, the SPEC interfaces and the corresponding surfaces (identified by toroidal flux) of a high radial resolution VMEC equilibrium are shown for a low $N_V$ case in \Fig{approximationtosmoothPoincare}.
  Despite the fact that a smooth profile was supplied to VMEC but a stepped-pressure profile was supplied to SPEC, the agreement is reasonable.
  This may be expected: the Shafranov shift depends \cite{White_01} primarily on the {\em integral} of the pressure profile and not, to lowest order, on the precise details of the pressure profile itself.
  To quantify the discrepancy, we compare the location of the magnetic axes.
  As shown in \Fig{axisconvergence}, as $N_V$ is increased the agreement improves.
  The stepped pressure profile approximation and the smooth pressure profile are shown in \Fig{steppedsmooth}.

  Indeed, it may be shown that the infinite interface limit of MRXMHD (i.e. assuming continuously nested magnetic flux surfaces) reduces to \mbox{$\nabla p = {\bf j} \times {\bf B}$} and \mbox{${\bf j}\cdot {\bf n}=0$}, where ${\bf n}$ is normal to the flux surfaces.
  These are the equations that define ideal MHD equilibria.
  The details of this derivation will be presented in a forthcoming paper.

  To illustrate a non-axisymmetric global equilibrium, we return to the `perturbed' equilibrium shown in \Fig{BeltramiPoincare}.
  Assuming that the radial sub-grid resolution is sufficiently high so that the error is dominated by finite Fourier resolution, we now confirm that the error in the interface geometry decreases as $M$ and $N$ increase.

  Let the exact solution for the $l$-th interface be described by $R(\theta,\zeta)$ and $Z(\theta,\zeta)$.
  The error between this and an approximation, $R_{M,N}(\alpha,\z)$ and $Z_{M,N}(\alpha,\z)$, with a potentially different poloidal angle, $\alpha$, on the $\zeta_0$ plane is quantified by $\Delta \equiv \int D(\theta) \, dl$, where $dl^2 \equiv dx^2+dy^2$ is the arc-length of the curve \mbox{$x(\theta) \equiv R(\theta,\zeta_0)$} and \mbox{$y(\theta)\equiv Z(\theta,\zeta_0)$}, and $D$ is the distance between this reference curve and the curve described by \mbox{$x_{M,N}(\alpha)\equiv R_{M,N}(\alpha,\zeta_0)$} and  \mbox{$y_{M,N}(\alpha)\equiv Z_{M,N}(\alpha,\zeta_0)$}, i.e. $D^2 \equiv \left[ x(\theta)-x_{M,N}(\alpha) \right]^2 + \left[ y(\theta)-y_{M,N}(\alpha) \right]^2$.
  The comparison is made at the same polar angle, so that \mbox{$y(\t) / [x(\t)-x_0] = y_{M,N}(\alpha) / [x_{M,N}(\alpha)-x_0]$}, where $x_0$ is a reference point (e.g. the magnetic axis).

  \insertsglfigure{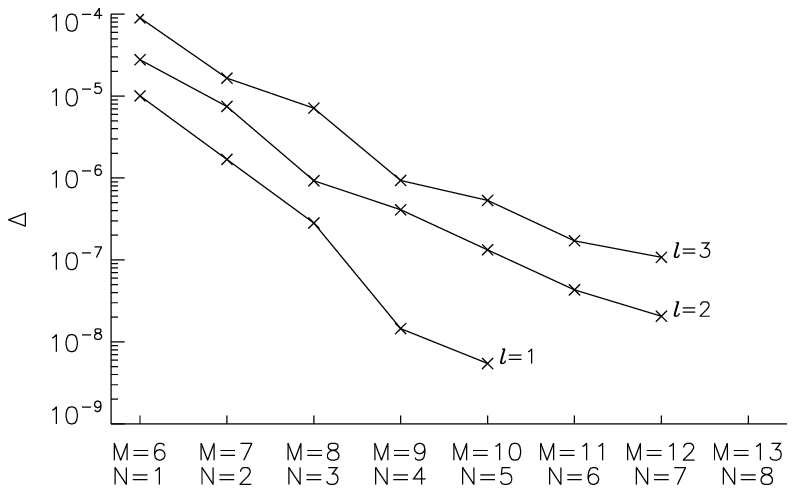}{Difference between finite $M$, $N$ approximation to interface geometry, and a high-resolution reference approximation (with $M=13$ and $N=8$), plotted against Fourier resolution.}{FourierConvergence}

  As the exact solution is not known apriori, we take as the reference configuration the highest resolution approximation available.
  The error in the interface geometry, for each of the internal interfaces, is shown as a function of $(M,N)$ in \Fig{FourierConvergence}, where we see that the error decreases as the Fourier resolution is increased.

  All properties of the equilibrium are defined by the interface geometries: if the interface geometry has converged, then so too have the Beltrami fields, and the location and size of the magnetic islands and chaotic seas -- well, it is difficult to resolve the infinitely complicated structure of phase space, but ${\bf B}$ itself and all integral properties are converged.

  The good convergence properties of the interface geometry with Fourier resolution is because (i) the interfaces are chosen to have the most noble transform, and so are `as far away as possible', so to speak, from the lowest order islands and associated chaos, and so are the smoothest surfaces (this is in contrast to flux surfaces adjacent to a separatrix, or flux surfaces that are nearly critical); and (ii) that the Fourier representation exploits a preferred angle parametrization that minimizes the spectral width.
  More importantly, perhaps, is that convergence is obtained because there is a well defined, exact solution, and that the numerical discretization is capable of resolving all the structure of the solution. 

  As a final illustration, we present a stepped-pressure equilibrium consistent with the boundary and profiles obtained via a 3D STELLOPT reconstruction \cite{Lazerson_12} of an up-down symmetric DIIID experimental shot with applied resonant magnetic perturbation (RMP) fields.
  The reconstruction process seeks to infer the experimental configuration by adjusting the MHD equilibrium (presently, STELLOPT is built around VMEC) by varying the the plasma boundary and the pressure and current profiles, so that derived quantities (such as Thomson scattering, motional Stark effect polarimetry, and magnetic diagnostics) match the experimental measurements.
  Because of the applied error fields, and the plasma response to these error fields, the reconstructed boundary is slightly, but significantly, perturbed from axisymmetry.
  The pressure and $q$-profiles, where $q$ is the safety-factor $q\equiv 1/\iotabar$, derived from the reconstruction are shown in \Fig{DIIIDpressure}.
  It is interesting to observe that the reconstructed pressure profile appears quite flat across the lowest order rational surfaces.
  Furthermore, the locations of locally maximum pressure gradient appear to coincide with strongly-irrational surfaces.

  We compute the stepped-pressure equilibrium using the reconstructed boundary, a stepped, $N_V=32$ approximation (\Fig{DIIIDpressure}) to the reconstructed pressure profile, and the reconstructed $q$-profile.
  The rotational transforms of the interfaces are chosen by selecting the most noble irrationals that are within range.
  The Fourier resolution is $M=10$ and $N=6$, and the total radial sub-grid resolution is $279$.
  A \Poincare plot is shown in \Fig{DIIIDPoincare}; most visible is a $q=2$ island at where VMEC has the $q=2$ rational surface.
  In this `fixed-boundary' calculation, the boundary is constrained to remain a fixed, good flux surface.
  To determine to what extent the RMP fields and the plasma response ergodize the field in the vicinity of the plasma edge, a `free-boundary' calculation is required.
  This is left for future work.

  \insertsglfigure{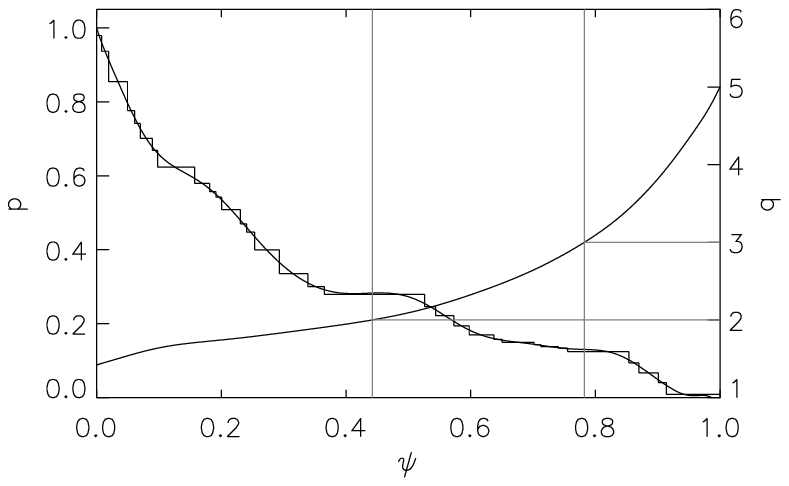}{Pressure profile (smooth) from a DIIID reconstruction using STELLOPT and stepped-pressure approximation. Also shown is the inverse rotational transform $\equiv$ safety factor.}{DIIIDpressure}
  \insertsglfigure{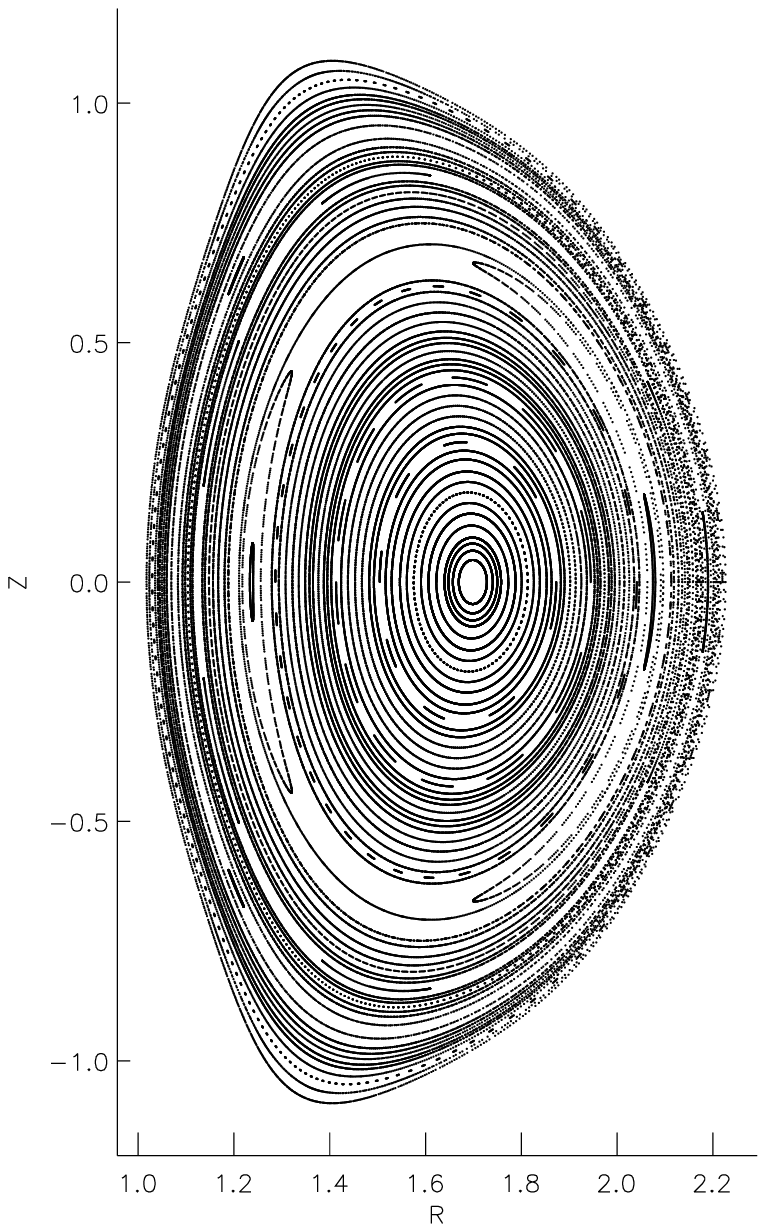}{\Poincare plot of a DIIID equilibrium with perturbed boundary, calculated using SPEC.}{DIIIDPoincare}

 \section{comments, future Work}

  The fact that stepped-pressure equilibria can be derived as minima of an energy functional is a great convenience numerically, as this allows employment of minimization methods to construct the Beltrami fields and interface geometries that satisfy force balance.
  Furthermore, the MRXMHD energy functional provides a self-consistent approach for determining the stability of partially chaotic equilibria \cite{Hole_Hudson_Dewar_07,Mills_Hole_Dewar_09,HMHD_09}.
  In future work, we hope to explore whether the suppression of edge localized modes by resonant magnetic perturbations \cite{Evans_Moyer_Thomas_etal_04}, that result in a stochastic plasma edge, may be understood through an MRXMHD stability analysis.
 
  In the above text we have distinguished MRXMHD equilibria, for which the algorithm allows $\mu$ to vary to force the helicity constraint, from stepped-pressure equilibria, for which the algorithm allows both $\mu$ and the poloidal flux to vary to force the transform constraint and for which the self-consistent helicity is computed aposteori.
  The distinction is, however, superficial.
  The profiles could also be specified by keeping $\mu$ and the poloidal flux as fixed input parameters.
  That the profiles can be prescribed and constrained in a variety of ways illustrates the flexibility of the theoretical and numerical method.
  (Note that because ideal MHD is not applied globally, there is no explicit relationship between the toroidal and poloidal fluxes and the rotational-transform profile, for example.)
  To decide how the profiles should be described, one must depart from equilibrium theory and develop a self-consistent model of transport, which may suggest how, for example, the poloidal flux, the parallel current and the helicity should vary in time to preserve the transform constraint.
 
  Given that the MRXMHD equilibria do not have $1/x$ and $\delta$-function currents at the rational surfaces, and that, in a pressure transport model that includes a small perpendicular diffusion, maximum pressure gradients will appear at the most irrational locations \cite{Hudson_Breslau_08,Hudson_09} -- and so too will, perhaps, the pressure driven currents -- then it would seem that MRXMHD equilibria are smoothly connected (e.g. as the perpendicular-diffusion coefficient $\kappa_\perp \rightarrow 0$) to nearly-ideal steady-state equilibria, and so it may be advantageous to initialize resistive MHD initial value codes such as NIMROD \cite{NIMROD} or M3DC1 \cite{M3DC1} with SPEC equilibria.
  Local flattening of the pressure gradient across the resonances can provide increased stability and can allow access to higher plasma beta \cite{IWUTC_01,Ichiguchi_Carreras_11}.

  The equilibrium model considered in this article does not include plasma flow, which can impact both the equilibrium and island healing phenomena \cite{CCH_11,CCH_12}. 
  Extended MHD modeling of plasmas \cite{IWUTC_01,BCC_10,FJS_10,Ichiguchi_Carreras_11,SHSKH_12} is crucial for understanding some key experimental observations.
  Perhaps, by extending a stepped-pressure equilibrium, an equilibrium with a continuous-pressure profile could be constructed by replacing the pressure jump interfaces with finite-width ideal-MHD layers \cite{Mills_Hole_Dewar_09}, each topologically constrained to avoid resonances and to avoid ideal instabilities \cite{Mills_Hole_Dewar_09}, and, perhaps, capable of supporting extended MHD behavior.
  A variational principle based on minimizing the generalized enstrophy \cite{Yoshida_Mahajan_02}, $ F = \int \left| \nabla \times \left( {\bf V}+{\bf A} \right) \right|^2 dx$, may better describe self-organization in two-fluid plasmas.

  In addition to the incremental code improvements that are inevitably required, the following in particular will be pursued.
  In this paper it was assumed that the plasma boundary, $\partial \cal V$, is a given, {\em fixed} toroidal surface.
  More generally, the magnetic field is part generated by external currents, and the {\em free}-boundary problem could be solved with a little extra work, where $\partial{ \cal V}$ is to be determined as part of the solution.
  Also, most modern tokamaks are not up-down symmetric, so it will be required to relax the stellarator symmetry constraint.

  It is worth exploring more efficient numerical methods for computing the Beltrami fields.
  In a closed domain $\cal P$ in ${\mathbb R}^3$, in general multiply connected, the solutions of \mbox{$\nabla \times {\bf B} = \mu {\bf B}$} can be represented \cite{Kress_81,Kress_86} by
 \begin{eqnarray}
 \vec{B} = (\curl\, + \mu)\!\!\int_{\partial\cal P}\!\! G(\vec{r},\vec{r}')\,\vec{B}'\times\vec{n}'\, \d S' \;,
 \label{eq:GreensFnSoln}
 \end{eqnarray}
 where $\vec{n}$ is the outward unit normal on $\partial\cal P$ and $G(\vec{r},\vec{r}')$ 
       satisfies $(\nabla^2 + \mu^2)G(\vec{r},\vec{r}') = -\delta(\vec{r} - \vec{r}')$, with $\delta(\cdot)$ being the 3D Dirac $\delta$ function.

  Given the geometry of an interface, the maximum pressure jump that an interface can support can be quickly determined by an analysis of the pressure-jump Hamiltonian: the pressure discontinuity, $2(p_+-p_-)$, is increased until the appropriate irrational surface of the pressure-jump Hamiltonian is critical.
  Looking beyond our present task of constructing an equilibrium consistent with a given pressure that is not changed, this gives an efficient method for distributing the pressure so that the most robust interfaces support the most pressure.
  As the pressure across any interface is altered, there will be a global response that requires re-computation of the equilibrium.

  We dedicate this article to Paul Garabedian, whom we consider to be a pioneer in the field of 3D MHD calculations, and who also endorsed and employed weak solutions \cite{Garabedian_98}.
  We are also indebted to Steve Hirshman, who has provided VMEC.
  One of us (SRH) acknowledges stimulating discussions with Neil Pomphrey, with Don Monticello regarding the construction of straight-field-line coordinates, and with Allen Boozer regarding the choice of gauge \cite{Finn_Chacon_05} for the vector potential, and RLD acknowledges discussion with Robert MacKay.
  The authors gratefully acknowledge support of the U.S. Department of Energy and the Australian Research Council, through grants DP0452728, FT0991899 and DP110102881.

 \end{document}